\documentclass[12pt]{article}
\usepackage[usenames, dvipsnames]{xcolor}
\usepackage{jcapmod}

\usepackage{tocloft}
\usepackage[normalem]{ulem}
\usepackage[]{todonotes}
\usepackage{verbatim}
\usepackage{mathrsfs}
\usepackage{cleveref}
\usepackage[margin=.8in]{geometry}

\usepackage{multirow}

\usepackage{caption}
\usepackage{longtable}
\usepackage{afterpage}
\usepackage{tikz}
\usepackage{setspace,caption}
\usepackage{lipsum}
\usetikzlibrary{matrix,arrows,calc}
\usepackage{bbm}
\usepackage{slashed}
\usepackage{graphicx}
\usepackage{subcaption}
\usepackage{psfrag}
\usepackage{tensor}
\usepackage{fouridx}
\usepackage{bm}
\usepackage{mdframed}
\usepackage{multirow}
\usepackage{soul}
\usepackage{bbold}
\usepackage{multicol}
\usepackage{tikz-cd}
\usepackage{rotating}

\usetikzlibrary{arrows}

  \usepackage{mdframed}
\tikzset{
commutative diagrams/.cd,
arrow style=tikz,
diagrams={>=latex}}

\usepackage{amsmath}
\usepackage{amssymb}
\usepackage{mathtools}

\usepackage{feynmf}
\usepackage{marvosym}

\usepackage{import}

\usepackage[utf8]{inputenc}
\usepackage{amsmath}
\usepackage{tikz}
\usepackage{extpfeil}
\usepackage{amsfonts}
\usepackage{amssymb}
\usepackage{bm}
\usepackage{graphicx}
\usepackage{array}
\usepackage{lipsum}
\usepackage{float}
\usepackage{multirow}
\usepackage{hhline}
\usepackage{tabularx}
\usepackage{graphicx}
\usepackage{afterpage}
\usepackage{longtable}
\usepackage{epstopdf}
\usepackage{ytableau}
\usetikzlibrary{decorations.markings}

\usepackage[titletoc]{appendix}

\newcommand{\cO}{\mathcal{O}}

\newcommand{\jht}[1]{{}}
\newcommand{\jt}[1]{{}}

\definecolor{cobalt}{RGB}{44, 98, 120}
\definecolor{celadon}{rgb}{0.67, 0.88, 0.69}
\definecolor{dm}{cmyk}{.20, 0, .30, 0}
\definecolor{burgundy}{rgb}{0.5, 0.0, 0.13}
\definecolor{plotBlue}{RGB}{94, 130, 181}

\newcommand*\xoverline[2][0.75]{
    \sbox{\myboxA}{$\m@th#2$}
    \setbox\myboxB\null
    \ht\myboxB=\ht\myboxA
    \dp\myboxB=\dp\myboxA
    \wd\myboxB=#1\wd\myboxA
    \sbox\myboxB{$\m@th\overline{\copy\myboxB}$}
    \setlength\mylenA{\the\wd\myboxA}
    \addtolength\mylenA{-\the\wd\myboxB}
    \ifdim\wd\myboxB<\wd\myboxA
       \rlap{\hskip 0.5\mylenA\usebox\myboxB}{\usebox\myboxA}%
    \else
        \hskip -0.5\mylenA\rlap{\usebox\myboxA}{\hskip 0.5\mylenA\usebox\myboxB}%
    \fi}
\makeatother

\begin{document}

\newcommand{\main}{.}
\begin{titlepage}

\setcounter{page}{1} \baselineskip=15.5pt \thispagestyle{empty}

\bigskip\

\vspace{1cm}
\begin{center}
{\Large \bfseries
6D Anomaly-Free Matter Spectrum  \\ \vspace{.2cm} in F-theory on Singular Spaces}
\end{center}

\vspace{0.25cm}

\begin{center}
Antonella Grassi$^{1,4}$, James Halverson$^{2,6}$, Cody Long$^{3}$,\\ \vspace{.1cm} Julius L. Shaneson$^4$, Benjamin Sung$^2$, and Jiahua Tian$^5$ \\

\vspace{1 cm}
\emph{$^1$Dipartimento di Matematica, Università di Bologna \\ Bologna 40126, Italy}\\
\vspace{.3cm}
\emph{$^2$Department of Physics, Northeastern University \\ Boston, MA 02115, USA}\\
\vspace{.3cm}
\emph{$^3$Jefferson Physical Laboratory, Harvard University \\ Cambridge, MA 02138, USA}\\
\vspace{.3cm}
\emph{$^4$Department of Mathematics, University of Pennsylvania \\ Philadelphia, PA 19104, USA}\\
\vspace{.3cm}
\emph{$^5$The Abdus Salam International Centre for Theoretical Physics \\ Trieste 34151, Italy} \\
\vspace{.3cm}
\emph{$^6$The NSF AI Institute for Artificial Intelligence \\ and Fundamental Interactions}

\vspace{0.5cm}
\end{center}

\vspace{1cm}
\noindent

\begin{abstract}
	\ In this paper we study the 6d localized charged matter spectrum  of F-theory directly on  a singular elliptic Calabi-Yau 3-fold, i.e. without smoothing via resolution or deformation   of the entire fibration. Given only the base surface, discriminant locus, and the $SL(2,\mathbb{Z})$ local system, we propose a general prescription for determining the charged matter spectrum localized at intersections of seven-branes, using the technology of string junctions. More precisely, at each codimension-$2$ collision of seven-branes, we determine the local seven-brane content and compute the number of massless string junctions modulo the action of the $SL(2,\mathbb{Z})$ monodromy. We find agreement with the predicted results from $6$d anomaly cancellation in all cases considered. Examples include a generic Weierstrass model with arbitrary Kodaira fiber intersecting an $I_1$, as well as cases with jointly charged matter localized at intersections of non-abelian seven-branes.

\end{abstract}

\end{titlepage}

\clearpage

\tableofcontents
\newpage
\section{Introduction}


F-theory ~\cite{Vafa:1996xn, Morrison:1996pp} is a non-perturbative formulation of type IIB superstring theory that geometrizes seven-brane physics. Specifically, the axio-dilaton profile sourced by the seven-branes is encoded in an elliptically fibered Calabi-Yau variety $X\to B$.
The singularities of $X$ encode the structure of seven-branes, providing the geometric and topological data that is crucial for determining the degrees of freedom of the F-theory compactification and their low-energy physics.
Though unbroken non-abelian gauge symmetry on seven-branes requires singularities in the F-theory description, many analyses nevertheless smooth the variety. For instance, a series of blowups and small resolutions \cite{Bershadsky:1996nh, Katz:1996th, Morrison:2011mb, Katz:2011qp, Esole:2011sm, Marsano:2011hv, Lawrie:2012gg, Hayashi:2014kca, Klevers:2014bqa} may yield a smooth Calabi-Yau fourfold $X^\sharp$, where M-theory on $X^\sharp$ corresponds to the Coulomb branch of an associated 3d $\mathcal{N}= 2$ theory, obtained by compactification of the 4d theory on a circle. Another approach is to obtain a smooth Calabi-Yau fourfold $X^\flat$ by a complex structure deformation \cite{Gaberdiel:1997ud, DeWolfe:1998bi, DeWolfe:1998zf,Cvetic:2010rq, Cvetic:2011gp, Grassi:2013kha, Grassi:2014sda, Grassi:2014zxa, Grassi:2014ffa, Grassi:2016bhs}, which corresponds to a Higgsing of the 4d $\mathcal{N}=1$ gauge group associated to the 7-branes. Both techniques are indirect, however, as the smoothing moves the theory to a different phase. There is no guarantee that the physics of F-theory on $X$ is completely captured by the geometry and topology of $X^\sharp$ or $X^\flat$.

Instead, to understand compactifications with unbroken seven-brane gauge symmetry it is preferable to study that phase directly, i.e. via the singular geometry and topology of $X$ itself. Doing so requires the development of new mathematics, such as in the study of F-theory on singular spaces via matrix factorizations \cite{Collinucci:2014taa} or via string junctions \cite{Grassi:2018wfy}; the latter built on a mathematical theory of topological string junctions \cite{Grassi:2014ffa} developed in the case of a smooth variety.

In this work we continue to develop a theory of F-theory on singular spaces that utilizes string junctions. Indeed, our approach emphasizes the definition of F-theory as $SL(2,\mathbb{Z})$-equivariant type IIB supergravity coupled with background $(p,q)$ $7$-branes. From this perspective, the background spacetime is smooth and it is natural to establish an algorithm that generalizes the original counting of localized charged and uncharged matter at the intersections of $D7$-branes in terms of open string states. String junctions thus serve as a natural tool, and our proposal reduces precisely to the original counting in the case that all $(p,q)$ $7$-branes are mutually local, i.e. in the weakly coupled type IIB limit.

Specifically, we give a description of computing the localized charged matter spectrum at the intersection of two irreducible components of the discriminant locus. The main results of the paper are as follows, organized according to the section in which they appear. In Section \ref{sec:description}:
\begin{itemize}
\item
 Given a discriminant of the form $\Delta = z^n \tilde{\Delta}$, we obtain sets $S_G, S_R$ of vanishing cycles corresponding to the seven-brane content obtained from the local monodromy around $z^n = 0 $, and also from the local monodromy around $\tilde{\Delta} = 0$ restricted to the plane $z = \epsilon$.
 We count the number of root junctions associated to the sets $S_G$ and $S_R$, using methods from \cite{Grassi:2018wfy}.
 \item The threefold geometry induces a monodromy on the junctions that is crucial to compute the charged matter spectrum and match what is known from anomaly cancellation. Specifically, we compute the monodromy matrices $M_GM_R, M_G M_R^{-1}$ and their inverses, and identify asymptotic charges (and therefore junctions with those charges) that are related by monodromies generated from this set.
 \item  We argue that these monodromies coincide with a representation of the fundamental group of the complement of $\mathbb{C}^2$ by the union of two lines through the origin, and argue that this coincides with the physics via a Higgsing.
 \item
In the cases with matter charged under a product gauge group $G \times G'$, we use the same prescription by choosing $S_G, M_G$ from the set of seven-branes forming one gauge group factor $G$, and the residual data $S_R,M_R$ from $G'$ together with the other residual seven-branes.
 \end{itemize}
We study many examples in Section \ref{sec:examples}. Specifically:
\begin{itemize}
\item
We verify the above prescription in all cases with a generic tuning of a single Kodaira fiber intersecting with a residual $I_1$, assuming normal crossing.
\item
We verify the above prescription for cases with jointly charged matter for $III \times III, IV_s \times IV_s$, and $IV_s \times III$, again assuming normal crossing.
\end{itemize}
A subtlety regarding localized neutral hypermultiplets is studied in Section \ref{sec:neutralhypers}. Specifically,
\begin{itemize}
\item We exhibit an example of tunings of $III-I_1$ which yields the same number of localized charged hypermultiplets but different localized neutrals depending on the tuning. In particular, our entire prescription is completely independent of details of the tuning. A specific example is given at the end of the section.
\end{itemize}
\section{Vanishing Cycles and Minimal Normal Factorizations}\label{sec:reptheory}

Consider an elliptic fibration $\pi:X\rightarrow B$ over a complex algebraic surface $B$ described by a Weierstrass model
\begin{equation}
y^2 = x^3 + fx + g\,,
\end{equation}
where $f\in \mathcal{O}(-4K_B), g \in \mathcal{O}(-6K_B)$. The fiber degenerates along the discriminant locus $D \subset B$
\begin{equation}
D:=\{\Delta = 4f^3+27g^2 = 0\} \,,
\end{equation}
which is the location of the seven-branes.

\vspace{.2cm}
\noindent \textbf{Smooth case}

For the sake of simplifying the discusion, we first assume that the total space $X$ is non-singular and that all vanishing cycles are simple $(p,q)$-cycles, i.e., the vanishing cycle is $pa+qb$ where $a$ and $b$ are the meridian and longitudinal cycles of the elliptic curve $E$, respectively. We will refer to this as a simple type degeneration, where all codimension-one singularities are of type $I_1$. The $SL(2,\mathbf{Z})$ monodromy matrix associated with looping around a component of $D$ takes the form
\begin{align}
	M_{(p,q)}=\begin{pmatrix}
		1-pq & p^2 \\
		-q^2 & 1+pq
	\end{pmatrix}.
\end{align}

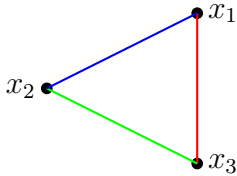
\begin{figure}[h]
	\centering
	\begin{tikzpicture}
		\filldraw[black] (-1,0) circle (2pt) node[anchor=east] {$x_2$};
		\filldraw[black] (1,1) circle (2pt) node[anchor=west] {$x_1$};
		\filldraw[black] (1,-1) circle (2pt) node[anchor=west] {$x_3$};
		\draw[red, thick] (1,1) -- (1,-1);
		\draw[green, thick] (1,-1) -- (-1,0);
		\draw[blue, thick] (1,1) -- (-1,0);
	\end{tikzpicture}\caption{The three solutions $x_1,x_2,x_3$ to $x^3+fx+g=0$ are plotted at a generic smooth point. We define the cycles $\pi_i$ such that the red line segment denotes $\pi_1$, the green line segment denotes $\pi_2$ and the blue line segment denotes $\pi_3$.}
	\label{fig:xplane}
\end{figure}

In this work the correspondence between the geometry of the elliptic curve $E$ and its $SL(2,\mathbb{Z})$ representation, or the correspondence between geometric data and algebraic data, will play a crucial role. The geometric data can be read off by studying the motion of the three roots of the equation
\begin{equation}\label{eqn:rhs}
x^3 + fx + g = 0\, ,
\end{equation}
 as one completes a closed path encircling a component $D_i \subset D$. At a generic point $p$ away from $D$ the three roots are distinct, and we can choose a labeling of these roots $x_1,x_2,x_3$, which provides a canonical definition of the cycles $\pi_1:=(1,0)$, $\pi_2:=(-1,-1)$ and $\pi_3:=(0,1)$ given such a labeling. In Fig.~\ref{fig:xplane} we plot the roots in the $x$-plane, as well as the cycles. Such a choice is made up to a global $SL(2,Z)$ transformation. These labels denote the vanishing cycles as two roots collide along components of $D$.

The cycles $\pi_1$, $\pi_2$ and $\pi_3$ are a (over-complete) set of generators of $H_1 (E,\mathbb{Z})$, where $E$ is the fiber at $p$. 
Fixing a reference point $p$ away from $D$, as we approach $D$ a $(p,q)$-cycle vanishes, as we are only considering degenerations of a simple type, i.e., we are considering smooth $X$ with only $I_1$ fibers. For instance, if a $(1,0)$-cycle vanishes, $x_1$ and $x_3$ approach each other and become degenerate on $D$. To properly analyze the monodromy, one should then perform a loop around $D$; for instance, taking a loop around $D$, for which the vanishing cycle is $(1,0)$, will induce a geometric swap of the roots $x_1$ and $x_3$.

\vspace{.2cm}
\noindent \textbf{Singular case}

However, the case that the total space $X$ is singular is both more physically interesting, and more generic, and we can no longer restrict ourselves to simple type degenerations. When approaching $D$, $x_1$, $x_2$ and $x_3$ will usually all degenerate into one point. In such cases one needs further tools to analyze the vanishing cycles and monodromy. As discussed in the Introduction, this may be done either via resolution or deformation, but both methods are indirect (likely losing information of the unbroken phase), and we prefer to work directly with the singular space. Furthermore, in nearly all known cases (e.g. \cite{Taylor:2015ppa, Halverson:2017ffz}) a smoothing complex structure deformation does not exist. Similarly, some singularities that give rise to localized neutral hypermultiplets cannot be resolved \cite{Arras:2016evy}.


When $X$ is singular, $D$ is in general a non-reduced scheme, where each component's multiplicity corresponds to the number of $(p,q)$ 7-branes along that component. In this case one cannot read off the vanishing cycles by approaching the components $D_i$ of $D$ (from an appropriately chosen fixed point), but instead must infer a set of vanishing cycles via the motion of the roots induced by traversing a loop around each $D_i$. While the motion of the roots is in general not a simple exchange (except in the simple $I_1$ case), any motion can be decomposed into an ordered set of exchanges. Such an ordered set can be used to define a set of vanishing cycles. This choice corresponds to a decomposition of the monodromy matrix $M$ into a minimal normal factorization (MNF), defined in \cite{velez2008normal}. An MNF of a monodromy matrix is a decomposition of an $SL(2,\mathbb{Z})$ monodromy matrix associated with a Kodaira fiber type into $n$ factors with each factor given by one of the following matrices corresponding to the monodromy sourced by a $(1,0)$ and a $(0,1)$ seven-brane respectively.
\[
M_{(1,0)} = \begin{pmatrix}
		1 & 1\\
		0 & 1
	\end{pmatrix},	
\qquad
M_{(0,1)} = \begin{pmatrix}
		1 & 0\\
		-1 & 1
	\end{pmatrix}
\]
Such a factorization exists for each $SL(2,\mathbb{Z})$-matrix corresponding to a Kodaira fiber and is unique up to Hurwitz moves and we refer to \cite[section 3.2]{Grassi:2018wfy} for an in-depth summary of the results of \cite{velez2008normal} and its implications for singular string junctions. A Hurwitz move is given by one of the following transformations for $g_i \in G$ (for our purpose, $G = SL(2,\mathbb{Z})$)
\begin{align*}
	g_1g_2\cdots g_ig_{i+1}\cdots g_k\rightarrow g_1g_2\cdots g_{i+1}(g_{i+1}^{-1}g_ig_{i+1})\cdots g_k
\end{align*}
or
\begin{align*}
	g_1g_2\cdots g_ig_{i+1}\cdots g_k\rightarrow g_1g_2\cdots (g_ig_{i+1}g_i^{-1})g_i\cdots g_k,
\end{align*}
i.e., $g_{i+1}$ is ``pulled past'' $g_i$, conjugating it in the process, or vice versa.

Given a disc $C \subset B$ intersecting components of the discriminant locus, we may consider the restriction to an elliptic surface $X_1 \rightarrow C$. In the following, we will use the technology of string junctions assuming that they exist as a basis of $H_2(\tilde{X}_1,E_p)$ where $\tilde{X}_1$ is the total space of a deformation of $X_1 \rightarrow C$ to an elliptic surface with only $I_1$ singularities. We will use a canonical ordering of the vanishing cycles (MNF) given in Table~\ref{tab:summary} and the associated intersection pairing, but the results will be independent of the choice of basis. Despite such assumptions, we emphasize that all these properties of string junctions can be obtained completely algebraically independent of the deformation and in \cite{bcjt}, we will demonstrate that all the relevant data such as the intersection pairing are indeed invariant under Hurwitz moves.

\section{Matter in 6D F-theory  Compactifications on Singular Spaces}\label{sec:description}

Given a Weierstrass or Tate model of an elliptically fibered Calabi-Yau threefold with a gauge group supported on a divisor $D_z := \{ z= 0\}$ with $z$ a local coordinate on $B$, the discriminant locus takes the form
\begin{align}
	\Delta = z^n\tilde\Delta\, ,
\end{align}
where $\tilde\Delta$ is the residual discriminant locus. Generally there are matter hypermultiplets localized at intersection points $z = \tilde\Delta = 0$, and the goal of this work is to count such hypermultiplets. At the intersection the fiber degenerates further, and the gauge algebra is enhanced to a larger one. 
To study the spectrum at the enhancement point we will probe the point with a D3-brane probe.

As discussed in the previous section we can associate to $D_z$ and (each component of) $\tilde\Delta$ an ordered set of vanishing cycles via an MNF. At the collision point the vanishing cycles intersect, and massless matter can be realized by closed string junctions with one boundary supported on $D_z$, and one boundary supported on $\tilde\Delta$. A direct counting of these string junction states supported at the codimension-two locus generally gives a matter spectrum that is too large to satisfy anomaly cancellation. However, we will propose that these states, labeled by their asymptotic charges with respect to $D_z$, are not all independent, and are related by a set of monodromy actions that identifies some states with one another. To see this, we will focus on a local intersection point of $D_z$ and $\tilde\Delta$, which we label $u$, and probe the enhancement point with a mobile D3 brane. The D3 brane can traverse loops in $B\setminus D$, which will induce an action on the string junctions extending from $D_z$, to be attached to $\tilde\Delta$. The relevant monodromies will be those associated to the point of enhancement, in the following way: fix a point $p$ in a local neighborhood of $u$. Let us for now assume that $D_z$ and $\tilde\Delta$ are locally normal crossing schemes, not necessarily reduced, and let the monodromy matrices $M_G$ and $M_R$ correspond to the monodromy action around $D_z$ and $\tilde\Delta$ respectively (G is for gauge and R is for residual) . We can then associate with $u$ the monodromy matrices $M_R \cdot M_G$, $M_G \cdot M_R$, $M_R \cdot M_G^{-1}$, $M_R^{-1} \cdot M_G$, and their inverses. Let us call this set of matrices $\{M_u^i\}$, $i = 1\dots 8$.

\begin{figure}[h]
	\centering
	\begin{tikzpicture}
		\draw[blue, thick] (-3,0) -- (3,0);
		\draw[red, thick] (0,2) -- (0,-2);
		\draw[black, thick] (-.8,1.2) -- (1.2,-.8);
		\node[draw] at (2,2) {$\epsilon \neq 0$};
		\filldraw[black] (0,0) circle (2pt) node[anchor=north east] {$u$};
	\end{tikzpicture}
	\quad \quad
 	\begin{tikzpicture}
		\draw[blue, thick] (-3,0) -- (3,0);
		\draw[red, thick] (0,2) -- (0,-2);
		\draw[black, thick] (-1,1) -- (1,-1);
		\node[draw] at (2,2) {$\epsilon = 0$};
		\filldraw[black] (0,0) circle (2pt) node[anchor=north east] {$u$};
	\end{tikzpicture}\caption{The gauge 7-branes are on top of each other and denoted by the blue line. The residual discriminant locus is denoted by the red line, which for the moment we assume to be normal crossing. The black line segment denotes a disk that intersects the 7-branes. In the left figure the disc misses the codimension-two point $u$, but intersects the codimension-one loci each at a point. In the right figure the disc has been deformed to intersect $u$ at a point. }
	\label{fig:general_geometry}
\end{figure}
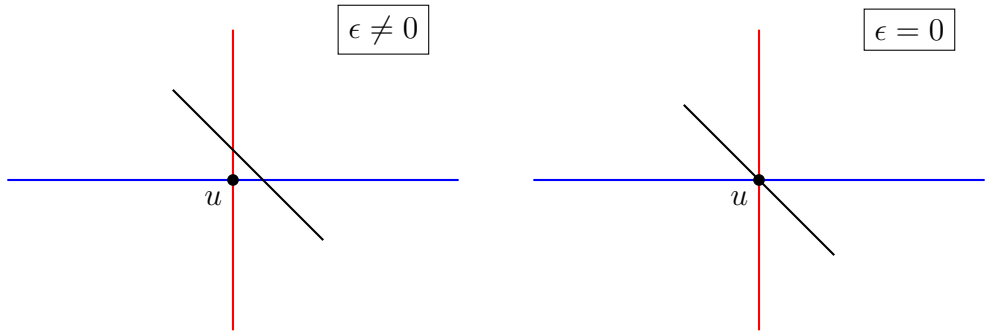

\begin{figure}[h]
		\centering
	\begin{tikzpicture}
		\filldraw[red] (-1.5,0) circle (2pt) node[anchor=north] {$\tilde\Delta = 0$};
		\filldraw[blue] (1.5,0) circle (2pt) node[anchor=north] {$z = 0$};
		\draw[thick] (0,0) ellipse (4cm and 2.5cm);
		\draw[dashed,decoration={markings, mark=at position 0.25 with {\arrow{>}}},postaction={decorate}] (-1.5,0) circle (1.5cm);
		\draw[dashed,decoration={markings, mark=at position 0.25 with {\arrow{>}}},postaction={decorate}] (1.5,0) circle (1.5cm);
		\node at (-1.5,1.1) {$L_R$};
		\node at (1.5,1.1) {$L_G$};
		\node[draw] at (4,2) {$\epsilon \neq 0$};
	\end{tikzpicture}
	\vspace{.5cm}
	\begin{tikzpicture}
		\filldraw[purple] (0,0) circle (2pt) node[anchor=north] {$u$};
		\draw[thick] (0,0) ellipse (4cm and 2.5cm);
		\draw[dashed,decoration={markings, mark=at position 0.25 with {\arrow{>}}},postaction={decorate}] (0,0) ellipse (3cm and 1.5cm);
		\node at (0,1.1) {$L = L_R \cdot L_G$};
		\node[draw] at (4,2) {$\epsilon = 0$};
	\end{tikzpicture}\caption{On the disk $D$ near $z = \tilde\Delta = 0$ the $D3$ probe can either traverse the loop $L = L_R\circ L_G$ or the loop $L^{-1} = L_G^{-1} \circ L_R^{-1}$. $L_G$ and $L_R$ are as directed by the arrows. This disc can be deformed to only contain the codimension-two point $u$, by taking $\epsilon \rightarrow 0$.}
	\label{fig:disk}
	\end{figure}
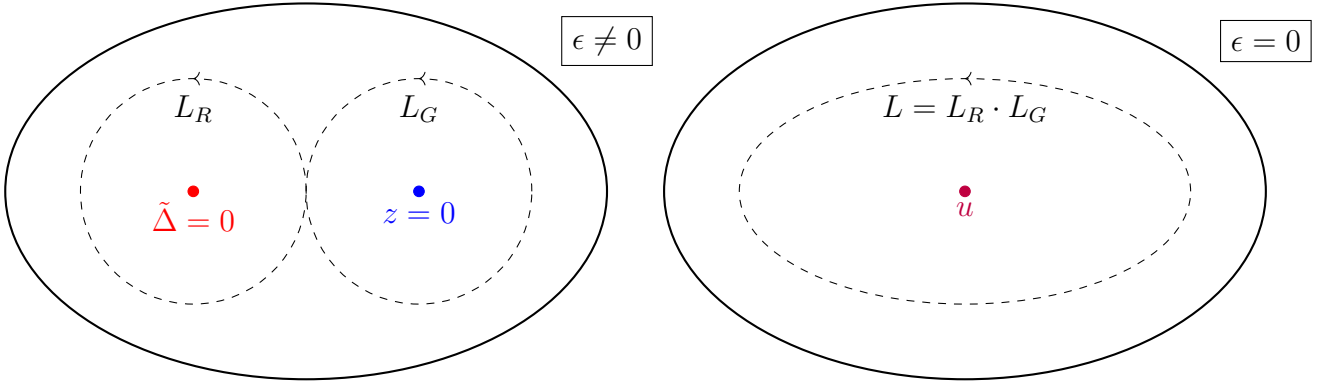

These matrices have a clear geometric interpretation, realized by a D3 probe approaching $u$. In order to probe $u$, we want to study the monodromy associated with $u$, and since $u$ is associated to the collision of two co-dimension one loci, we can naturally associate the above matrices as $u$-point monodromies.  Let us consider a point $p$ near $u$ on a disk that intersects both $z = 0$ and $\tilde\Delta = 0$, such that the distance on the disc between  $z = 0$ and $\tilde\Delta = 0$ is a small parameter $\epsilon$.\footnote{Distinguishing between a parameter distance and a proper distance will not be important here as all our studies occur at finite distances in moduli space.} For such a disc one can take the limit $\epsilon \rightarrow 0$, such that the only singularity on the disc is the codimension-two point $u$.    This is shown in Fig. \ref{fig:general_geometry}.
On the disk the geometry is sketched in Fig. \ref{fig:disk}. Clearly there are two generators of $\pi_1(B\backslash\Delta)$ on the disc, and the corresponding loops are $L_G$ and $L_R$, with corresponding monodromy matrices $M_G$ and $M_R$.   Taking the parameter $\epsilon \rightarrow 0$, we place the disc so that it only contains the point $u$ as a component of the discriminant $\Delta$. By taking the associated loops in the disc around $u$ we can probe the matter locus via monodromy. On such a disc it is clear there are two loops that can be deformed to encircle $u$ without coming into contact with the discriminant locus. 
These are $L = L_R\circ L_G$ and $L^{-1} = L_G^{-1} \circ L_R^{-1}$, with associated monodromy matrices $M = M_R \circ M_G$ and  $M^{-1} =   M_G^{-1} \circ M_R^{-1}$, respectively.

We require that the matter states supported at $u$ are given by the junctions that are invariant under both $M$ and $M^{-1}$. These monodromies can be read off by studying the motion of the three roots $x_1,x_2,x_3$ of $y^2 = x^3 + fx + g = 0$ upon traversing the loops $L$ or $L^{-1}$ as in \cite{Grassi:2016bhs,Grassi:2018wfy}. The geometric motion of the roots can then be associated with an $SL(2,\mathbb{Z})$ monodromy matrix that acts explicitly on the asymptotic charge $a(J)$ of a string junction. More precisely, the connecting homomorphisms
\begin{equation}
	\partial: H_2(X_1,E_p) \to H_1(E_p) 
\end{equation}
carries a monodromy on the junctions to the monodromy on the asymptotic charges, where $X_1 $ is the restriction to $D$ of a deformation of $\pi: X \to B $ to a fibration that has only $I_1$-singularities when restricted to  the intersection of $B$ with all but a finite set of parallel hyperplanes. We assume that, on $D$, at $z = 0$ there is a stack of gauge 7-branes $S_G = \{\pi_1,\dots,\pi_i\}$ and $\tilde\Delta = 0$ consists of 7-branes $S_{R} = \{\pi_{i+1},\dots,\pi_n\}$. Naively the matters corresponding to the string junctions $J_{GR}$ stretch between $S_G$ and $S_R$ and can be obtained using the technique developed in \cite{Grassi:2013kha}. When the brane content $S_G+S_R$ is associated with a gauge algebra these junctions are simply given by the branching rule $G_{S_G+S_R}\rightarrow G_{S_G}$ where $G_S$ is the gauge algebra corresponding to the set $S$ of 7-branes. The junctions $J_{GR}$ typically give rise to more matters than are required by 6D anomaly cancellation. The monodromies $M$ and $M^{-1}$ map different junctions with different $a(J)$'s to each other and this reduction leads to the correct 6D matter spectrum.

\begin{figure}[h]
	\centering
	\begin{tikzpicture}
		\draw[blue, thick] (-1,0) -- (2,0);
		\draw[red, thick] (0,-1) -- (0,2);
		\draw[black, thick, dashed] (1.5,0) -- (0,1.5);
		\filldraw[black] (0,0) circle (2pt) node[anchor=north east] {$u$};
		\filldraw[green] (0.75,0.75) circle (2pt) node[anchor=south west] {$v$};
		\filldraw[black] (0,1.5) circle (2pt) node[anchor=east] {$A$};
		\filldraw[black] (1.5,0) circle (2pt) node[anchor=north] {$B$};
	\end{tikzpicture}
	\caption{The D3 brane at $v$ denoted by the green dot is placed on the 7-7 junction $AB$ stretched between two stacks of 7-branes intersecting at $u$. The 7-7 junction is represented by the dashed line and the two stacks of 7-branes are represented by the red and blue lines. The two 3-7 junctions are $Av$ and $vB$ with asymptotic charges $a(J)$ and $-a(J)$ respectively and clearly they combine into the full 7-7 junction $AB$.}
	\label{fig:D3_betwee_77}
\end{figure}
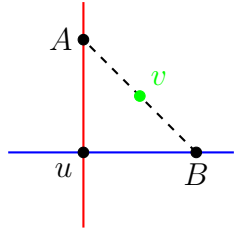

The physical interpretation is as follows: a massless state in a representation $\mathcal{R}$ under the gauge group $G$ corresponds to a string junction $J$ supported at $u$, with non-trivial asymptotic charge $a(J)$ on $D_z$. Such a junction pinches off on $D_z$, and also on $\tilde{\Delta}$, with the same asymptotic charge flowing into $\tilde{\Delta}$. We place a D3 probe on this junction, so that the matter state supported at $u$ can be broken into two junctions: a junction $J_G$ with one boundary on the D3 brane and the other on $D_z$, with asymptotic charge $a(J)$, and a junction $J_R$ with one boundary on the D3 brane and the other on $\tilde{\Delta}$, with asymptotic charge $-a(J)$, such that the junctions can join together to reproduce the full 7-7 junction, see Figure \ref{fig:D3_betwee_77}.
To probe these states we arrange for a disc $\mathcal{D}$ containing the point $u$ as the only point of $\Delta \cap D$, and then allow the D3 brane to loop around $u$. This will induce an action on the asymptotic charge of both $J_G$ and $J_R$, in a manner such that they can again be joined to create a junction corresponding to charged matter in $\mathcal{R}$ of $G$. The identification of such states will be necessary to produce the correct amount of charged matter to satisfy 6d anomaly cancellation.

As is clear from both the geometry and the pictures, the disc $\mathcal{D}$ is not unique. We require the disc, at $\epsilon \neq 0$, to intersect both $D_z$ and $\tilde{\Delta}$ at a point, and that these points collide as $\epsilon \rightarrow 0$. In fact, there are four such discs that locally achieve this, which correspond to the four quadrants in Fig.~\ref{fig:general_geometry}. Therefore, for each quadrant we can define the monodromy matrices $M$ and $M^{-1}$. Via the fact
\begin{align}\label{eq:geomrep}
	M_{L} = M_{L_2}\circ M_{L_1},\ \text{if}\ L = L_2\circ L_1\, ,
\end{align}
we have the total set of monodromy matrices associated to $u$: $M_R \cdot M_G$, $M_G \cdot M_R$, $M_R \cdot M_G^{-1}$, $M_R^{-1} \cdot M_G$, and their inverses which we above labeled $\{M_u^i\}$, $i = 1\dots 8$. By requiring the matter spectrum be invariant under all  $\{M_u^i\}$, we will find the correct amount of matter to satisfy 6d anomaly cancellation. We note that often many of these matrices will induce redundant identifications, and so in the examples we will only note the minimal set of non-trivial ones.



\begin{figure}[h]
	\centering
	\begin{tikzpicture}
		\draw[blue, thick] (-2.25,0) -- (2.25,0);
		\draw[red, thick] (0,2) -- (0,-2);
		\filldraw[black] (1,1) circle (2pt) node[anchor=west] {$p$};
		\draw[black, thick] (1,1) .. controls (.9,1.2) and (.5, 1.35) .. (.1,1.35);
		\draw[black, thick] (-.1,1.35) .. controls (-.5,1.35) and (-.9,1.2) .. (-1,1);
		\draw[->, black, thick] (-1,1) .. controls (-.9,.8) and (-.5,.65) .. (0,.65);
		\draw[black, thick] (0,.65) .. controls (.5,.65) and (.9,.8) .. (1,1);
		\draw[black, thick] (1,1) .. controls (.8,.9) and (.65,.5) .. (.65,0);
		\draw[black, thick] (.65,0) .. controls (.65,-.5) and (.8,-.9) .. (1,-1);
		\draw[->, black, thick] (1,-1) .. controls (1.2,-.9) and (1.35,-.5) .. (1.35,-.1);
		\draw[black, thick] (1.35,.1) .. controls (1.35,.5) and (1.2,.9) .. (1,1);
		\node[text width=.1cm] at (2.5,0) {};	
	\end{tikzpicture}
	\begin{tikzpicture}
		\draw[blue, thick] (-2.25,0) -- (2.25,0);
		\draw[red, thick] (0,2) -- (0,-2);
		\filldraw[black] (1,1) circle (2pt) node[anchor=west] {$p$};
		\draw[black, thick] (1,1) .. controls (.9,1.2) and (.5, 1.35) .. (.1,1.35);
		\draw[black, thick] (-.1,1.35) .. controls (-.5,1.35) and (-.9,1.2) .. (-1.04,1.04);
		\draw[black, thick] (-1.04,1.04) .. controls (-1.2,.9) and (-1.35,.5) .. (-1.35,.1);
		\draw[black, thick] (-1.35,-.1) .. controls (-1.35,-.5) and (-1.2,-.9) .. (-1,-1);
		\draw[->, black, thick] (-1,-1) .. controls (-.8,-.9) and (-.65,-.5) .. (-.65,0);
		\draw[black, thick] (-.65,0) .. controls (-.65,.5) and (-.8,.9) .. (-.9,.9);
		\draw[->, black, thick] (-.9,.9) .. controls (-.9,.8) and (-.5,.65) .. (0,.65);
		\draw[black, thick] (0,.65) .. controls (.5,.65) and (.9,.8) .. (1,1);
		 \node[text width=.25cm] at (2.55,0) {\large $\sim$};
		 \node[text width=0cm] at (.1,2) {\footnotesize $R$};
		 \node[text width=0cm] at (2,.25) {\footnotesize $G$};
		 \draw[teal, thick] (1,1) .. controls (.9, 1.3) and (.5, 1.5) .. (.1, 1.5);
		 \draw[teal, thick] (-.1,1.5) .. controls (-.5,1.5) and (-.9,1.3) .. (-1.15,1);
		 \draw[teal, thick] (-1.15,1) .. controls (-1.25,.9) and (-1.4,.5) .. (-1.4,.1);
		 \draw[teal, thick] (-1.4,-.1) .. controls (-1.4,-.5) and (-1.25,-.9) .. (-1.1,-1);
		 \draw[teal, thick] (-1.1,-1) .. controls (-.85,-.9) and (-.7,-.5) .. (-.7,0);
		 \draw[teal, thick] (-.7,0) .. controls (-.7,.5) and (-.85,.9) .. (-1,.95);
		 \draw[teal, thick] (-1,.95) .. controls (-.9,1.1) and (-.5,1.25) .. (-.1,1.25);
		 \draw[teal, thick] (.1,1.25) .. controls (.5,1.25) and (.9,1.1) .. (1,1);
		\node[text width=0cm] at (-1.5,1.4) {\color{teal} \footnotesize $L_G'^{-1}$};
	\end{tikzpicture}
	\begin{tikzpicture}
		\draw[blue, thick] (-2.25,0) -- (2.25,0);
		\draw[red, thick] (0,2) -- (0,-2);
		\filldraw[black] (1,1) circle (2pt) node[anchor=west] {$p$};
		\draw[black, thick] (1,1) .. controls (.9,1.2) and (.5, 1.35) .. (.1,1.35);
		\draw[black, thick] (-.1,1.35) .. controls (-.5,1.35) and (-.9,1.2) .. (-1,1);
		\draw[->, black, thick] (-1,1) .. controls (-.9,.8) and (-.5,.65) .. (0,.65);
		\draw[black, thick] (0,.65) .. controls (.5,.65) and (.9,.8) .. (1,1);
		\draw[black, thick] (1,1) .. controls (.8,.9) and (.65,.5) .. (.65,.1);
		\draw[black, thick] (.65,-.1) .. controls (.65,-.5) and (.8,-.9) .. (1,-1);
		\draw[->, black, thick] (1,-1) .. controls (1.2,-.9) and (1.35,-.5) .. (1.35,0);
		\draw[black, thick] (1.35,0) .. controls (1.35,.5) and (1.2,.9) .. (1,1);
		\node[text width=0cm] at (-1.1,1.5) {\footnotesize $L_R$};
		\node[text width=0cm] at (1.4,.3) {\footnotesize $L_G^{-1}$};
	\end{tikzpicture}
	\caption{Left two paths denote the possible paths of a D3 probe confined to a disk encircling the intersection of the two 7-brane stacks. The homotopy relation between the right two paths allows one to conclude the monodromy acting on the homology of the elliptic fiber.}
	\label{fig:monodromy}
\end{figure}
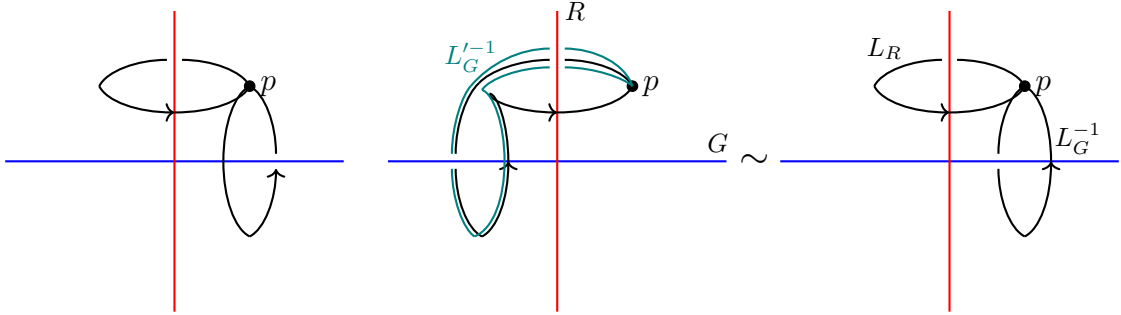

In the above discussion, we have associated natural monodromy matrices to distinguished paths encircling the codimension 2 point given by the two paths on the left of figure~\ref{fig:monodromy}. We now verify that the above assignments are in fact the correct monodromies realized by the desired paths. In particular, it will suffice to demonstrate that the second path $L_G'^{-1}\circ L_R$ in figure~\ref{fig:monodromy} is homotopy equivalent to the third path $L_G^{-1} \circ L_R$, or equivalently that $L_G'^{-1} \sim L_G^{-1}$. Under this assumption, the corresponding $SL(2,\mathbb{Z})$ monodromy representations are equivalent, and hence, all possible codimension 2 monodromies are generated by $M_G \cdot M_R$ and $M_G^{-1} \cdot M_R$.

To see the above homotopy, we will follow the discussion in \cite[Section~6.3]{MR2233951}. Denoting the axes by $G$ and $R$, the path $L_G$ is given by the circle in the plane $G = G_0$, $L_R$ by the circle in the plane $R = R_0$, and $p$ by the coordinates $(1,1)$. Consider the point $x(t)$ on the circle $R=R_0$ given by $(e^{it},1)$. The homotopy is given at each time $t$ by the path which begins at $p$, traverses to $x(t)$ along $R=R_0$, goes along the circle $(e^{it},e^{is})$ for fixed $t$, and then goes back to $p$. This induces the relation $L_G^{-1} \sim L_G'^{-1} \sim L_R \circ L_G^{-1} \circ L_R^{-1}$ and we conclude. In general, we note that the fundamental group of the complement of two normal crossing divisors in an open affine patch is the free abelian group of rank two, i.e. $\pi_1(\mathbb{C}^2 \setminus \{xy = 0\}, p) \cong \mathbb{Z}^2$.


\bigskip

So far we have assumed that the loci $\tilde{\Delta}$ and $D_z$ are normal crossing, in which case near $u$ it is clear that a disc $\mathcal{D}$ intersecting $D_z$ and $\tilde{\Delta}$ will capture all vanishing cycles at $u$. For general intersections between gauge group divisors and $I_1$-loci, this assumption will not be valid; for instance, in the type $III$-$I_1$ collision, the discriminant takes the form $\Delta = z^3(27z+4t^3)$, where at $u = \{ z= 0 \cap t = 0\}$, there is a multiplicity-three intersection in $t$, but away from $\{z = 0\}$ the $I_1$ locus splits into three components. In this case and similar ones, we must ensure that the $\epsilon \neq 0$ disc is large enough to intersect all components of the $I_1$ locus near $u$. Note that while $\tilde{\Delta}$ is in general one large connected subvariety, near $u$ it looks like three separate components, which intersect with multiplicity three at $u$, and we make $\mathcal{D}$ large enough to capture the three components, and thus the multiplicity three intersection as $\epsilon \rightarrow 0$.

We also note that, while the construction of the appropriate discs $\mathcal{D}$ is still achievable, one can instead (locally) start with a simpler model that is locally a normal crossing intersection, and then deform to the model of interest. Such a process corresponds to a Higgsing of a product gauge group $G^\prime \times G \rightarrow G$, where $G$ is the gauge group that we want to study. For example, in the $III -I_1$ case, we have
\begin{equation}
f = z F\, , g = z^2 G\, , \Delta = z^3 (4F^3 + 27 G^2 z)\,
\end{equation}
for a local coordinate $z$. Taking another local coordinate $F\sim t$, we have matter localized at $t = z = 0$. This is clearly not a normal crossing locus, but we can make it so, by taking $G = G^\prime t$, yielding
\begin{equation}
f = z t F^\prime\, , g = z^2 t^2 G^\prime\, , \Delta = z^3 t^3(4(F^\prime)^3 + 27 (G^\prime)^2)\, ,
\end{equation}
which is locally normal crossing, and corresponds to a $III-III$ intersection. From this normal crossing model we can arrange the appropriate disc near $u$, and then deform the geometry an infinitesimal amount to the geometry of interest.


The Higgsing described in the above paragraph then motivates the question: How does this prescription address the counting of localized massless matter for Weierstrass models with states charged under product representations $\mathcal{R}' \oplus \mathcal{R}$? Indeed, the obvious ambiguities in the naive extension of the aforementioned prescription to the case of intersections of multiple non-abelian 7-brane loci are what should we take $J_G$ to be and what is the corresponding codimension-$2$ monodromy. From the perspective of the above, there are also natural deformations to multiple distinct normal crossing loci.

In general, to count massless matter charged under the representation $(\mathbf{m},\mathbf{n})$ of the gauge group $G' \times G$, our prescription will be as follows. Without loss of generality, assume that the representation $\mathbf{n}$ of $G$ is not the trivial representation. We then choose a minimal deformation to a local normal crossing model $H\times G$ where $H$ corresponds to the 7-brane content of $G'$ and any residual 7-brane loci transverse to that of $G$. Such a deformation corresponds to a natural set of seven-branes $S = S_G \cup S_H$ and the gauge seven-branes within $S_H$ induces an additional partition $S_H = S_{G'} \cup S_R$. Given such partitions, a massless state in the representation $(\mathbf{m},\mathbf{n})$ then corresponds to a junction $J$ supported on $S$ and the prescription proceeds exactly as in the above with $J_G$ the sub-junction supported on $S_G$ and codimension-$2$ monodromy determined by the normal crossing model. In particular, if both factors in the product representation are non-trivial, this determines distinct deformations which yield the same counting of massless matter.

\vspace{.2cm}
We will illustrate the natural extension of our prescription for the $IV_s - III$ case, which exhibits an unusual feature: a forced codimension-$2$ collision of three $7$-branes, a phenomenon classified in \cite{Halverson:2016vwx}. In this example, we have the assignments
\begin{equation}
f = zt^{2}\, , g = z^2 t^2\, , \Delta = z^3 t^4(4t^2 + 27z )
\end{equation}
for $z$ and $t$ local coordinates. In particular, there is an additional $I_1$ loci given by the residual discriminant intersecting the $SU(3) \times SU(2)$ point which is characteristic of this model. For each product representation of $SU(3) \times SU(2)$, we will deform to a local normal crossing model in the spirit of the above example. For massless matter charged under the representation $(\mathbf{m},\mathbf{n})$ of $SU(3) \times SU(2)$ with $\mathbf{m}$ non-trivial, we will look at the $t$-slice and for $\mathbf{n}$ non-trivial, we will look at the $z$-slice. For each slice, there is a corresponding 7-brane stack, a partition into two sets of gauge seven-branes and residual seven-branes, and a monodromy action.


Concretely, to compute the spectrum of massless matter charged under $(\mathbf{3},\mathbf{1})$, we consider the set of all junctions with support on the set of seven-branes obtained by setting $t = \epsilon$. Given the partition of seven-branes $S_t = \{ S_{IV_s}\, | \,S_{III}\, | \,I_{1} \}$, we determine the junctions $J$ charged under the $\mathbf{3}$ of $SU(3)$ given by the gauge branes $S_{IV_s}$ and uncharged under the gauge branes $S_{III}$. We then compute the monodromy orbit of the asymptotic charge $a(J_{S_{IV_s}})$ under the codimension-$2$ monodromy determined by $S_t$ and identify the corresponding junctions. Likewise, we carry out a similar procedure for $(\mathbf{1},\mathbf{2})$ by setting $z = \epsilon$ and looking at the brane system $S_z = \{ S_{IV_s}\, | \,S_{III}\, | \,I_{1}\, | \,I_{1} \}$ and for the bi-fundamental representation $(\mathbf{3},\mathbf{2})$, both models yield identical results. This computation is detailed in Section \ref{sec:IVs*III}.

For most examples the monodromy action described thus far will be enough to reduce the spectrum to that required by anomaly cancellation. In fact, this works precisely when the to-be-identified charged matter states have different asymptotic charges, as the identification is done on the level of asymptotic charge. However, in the case where the over-counting is due to states with the same asymptotic charge there will be a further reduction. Consider a closed string $J$ that encircles $u$, such that $a(J)$ is an eigenvector of the $\{M_u^i\}$, with eigenvalue one. The lift of this closed string to $X_1$ is a cylinder with asymptotic charge $a(J) = 0$, and hence is an element in $H_2(X_1,\mathbb{Z})$.  Since the monodromy matrices of $u$ act trivially on the junction, via a Hanany-Witten move one can pass the closed string directly through $u$, without the junction gaining prongs localized at $u$. In a local neighborhood of $u$, such a junction may appear non-trivial, as it can attach simultaneously to $\tilde{\Delta}$ and $D_z$, but upon taking the junction to $u$, where the candidate state would become massless, it becomes trivial, and therefore does not contribute to the matter spectrum. In terms of the monodromy matrices, the condition for such a state to be trivial is
\begin{equation}
(M_u^i - \mathbb{1})\cdot a(J) = 0\, ,
\end{equation}
for all $i$. We will refer to such states as HW-trivial. We will find that the combination of the two monodromy actions, from the $u$-monodromy and the HW-triviality reduction, will reproduce the correct matter spectrum to satisfy anomaly cancellation.

Finally, we remark that our identification of junctions proceeds purely at the level of the asymptotic charges. Such an identification may or may not descend to an identification of the junctions themselves, and we will explore this in future work.
%
\\

We will use the following notation throughout:
\begin{align}
	\pi_1 := (1,0),\ \pi_2:= (-1,-1),\ \pi_3 := (0,1),\ \pi_\alpha := (1,-1),\ \pi_\beta := (2,1).
\end{align}
We will see the last two appear in the type $I_{n}$ fiber enhancement.

\section{Examples: Matter on Singular Spaces and Anomaly Cancellation}\label{sec:examples}

Having introduced a formalism for counting charged hypermultiplets in 6d F-theory compactifications on singular spaces, including a monodromy quotient necessary for obtaining the correct spectrum, we now apply it in many examples. We find that the matter spectra that we directly compute matches expectations from anomaly cancellation \cite{Grassi:2011hq, Morrison:2011mb, GrassiWeigand2}.

The data of all of the examples we will discuss in this section are summarized in Table \ref{tab:summary}.


\begin{table}[h]
\centering
\begin{tabular}{c|c|c|c|c|c}
	Fibration & Matter & Branes & $M_L$ & $M_{\overline{L}}$ & $N$ \\
	 \hline
	 \multirow{2}{*}{$I_n$} & \textbf{fund} & $\{\underline{\pi_1,\dots,\pi_1},\pi_1\}$ & $\begin{pmatrix}
	 	1 & n+1 \\
	 	0 & 1
	 \end{pmatrix}$ & $\begin{pmatrix}
	 	1 & n-1 \\
	 	0 & 1
	 \end{pmatrix}$ & 1 \\
	 \cline{2-6}
	  & $\mathbf{\Lambda^2}$ & $\{\underline{\pi_1,\dots,\pi_1},\pi_3,\pi_\beta,\pi_\alpha,\pi_2\}$ & $\begin{pmatrix}
	 	1 & n-8 \\
	 	0 & 1
	 \end{pmatrix}$ & $\begin{pmatrix}
	 	1 & n+8 \\
	 	0 & 1
	 \end{pmatrix}$ & 1 \\
	 \hline
	 $I_n^*$ & \textbf{vect} & $\{\underline{\pi_1,\pi_3,\pi_1,\pi_3,\pi_1,\pi_3,\pi_1,\cdots,\pi_1},\pi_1,\pi_1\}$ & $\begin{pmatrix}
	 	-1 & n-2 \\
	 	0 & -1
	 \end{pmatrix}$ & $\begin{pmatrix}
	 	-1 & n+2 \\
	 	0 & -1
	 \end{pmatrix}$ & 1 \\
	 \hline
	 $III$ & $\mathbf{2}$ & $\{\underline{\pi_1,\pi_3,\pi_2},\pi_1,\pi_3,\pi_2\}$ & $\begin{pmatrix}
	 	-1 & 0 \\
	 	0 & -1
	 \end{pmatrix}$ & $\text{Id}_{2\times 2}$ & 2 \\
	 \hline
	 $IV_s$ & $\mathbf{3}$ & $\{\underline{\pi_1,\pi_3,\pi_1,\pi_3},\pi_1,\pi_3,\pi_1,\pi_3\}$ & $\begin{pmatrix}
	 	0 & -1 \\
	 	1 & -1
	 \end{pmatrix}$ & $\text{Id}_{2\times 2}$ & 3 \\
	 \hline
	 $IV_s^*$ & $\mathbf{27}$ & $\{\underline{\pi_1,\pi_3,\pi_1,\pi_3,\pi_1,\pi_3,\pi_1,\pi_3},\pi_1,\pi_3,\pi_1,\pi_3\}$ & $\text{Id}_{2\times 2}$ & $\begin{pmatrix}
	 	-1 & 1 \\
	 	-1 & 0
	 \end{pmatrix}$ & 1 \\
	 \hline
	 $III^*$ & $\mathbf{56}$ & $\{\underline{\pi_1,\pi_3,\pi_1,\pi_3,\pi_1,\pi_3,\pi_1,\pi_3,\pi_1},\pi_3,\pi_1,\pi_3\}$ & $\text{Id}_{2\times 2}$ & $\begin{pmatrix}
	 	-1 & 0 \\
	 	0 & -1
	 \end{pmatrix}$ & $\frac{1}{2}$ \\
	 \hline
	 $III\times III$ & $(\mathbf{2},\mathbf{2})$ & $\{\underline{\pi_1,\pi_3,\pi_2},\underline{\pi_1,\pi_3,\pi_2}\}$ & $\begin{pmatrix}
	 	-1 & 0 \\
	 	0 & -1
	 \end{pmatrix}$ & $\text{Id}_{2\times 2}$ & 1 \\
	 \hline
	 $IV_s\times IV_s$ & $(\mathbf{3},\mathbf{3})$ & $\{\underline{\pi_1,\pi_3,\pi_1,\pi_3},\underline{\pi_1,\pi_3,\pi_1,\pi_3}\}$ & $\begin{pmatrix}
	 	0 & -1 \\
	 	1 & -1
	 \end{pmatrix}$ & $\text{Id}_{2\times 2}$ & 1 \\
	 \hline
	 \multirow{4}{*}{$IV_s\times III$} & \multirow{4}{*}{$\mathbf{R}$} & $t:\{\underline{\pi_1,\pi_3,\pi_1,\pi_3},\underline{\pi_1,\pi_3,\pi_1},\pi_3\}$ & $\begin{pmatrix}
	 	0 & -1 \\
	 	1 & -1
	 \end{pmatrix}$ & $\text{Id}_{2\times 2}$ & \multirow{4}{*}{1} \\
	 \cline{3-5}
	  & & $z:\{\underline{\pi_1,\pi_3,\pi_1},\underline{\pi_1,\pi_3,\pi_1,\pi_3},\pi_1,\pi_3\}$ & $\begin{pmatrix}
	 	0 & -1 \\
	 	1 & 0
	 \end{pmatrix}$ & $\begin{pmatrix}
	 	0 & 1 \\
	 	-1 & 0
	 \end{pmatrix}$ &
\end{tabular}
\caption{Summary of results. In each case we list the branes intersecting at the codimension 2 locus where the charged matters are localized and the monodromies corresponding to the branes. $N$ is the multiplicity of the matter in the representation as listed in the second column of the table. The branes that carry the gauge algebra are underlined in the third column of the table. In the $IV_s\times III$ case $\mathbf{R} = (\mathbf{3},\mathbf{2}) + (\mathbf{3},\mathbf{1}) + (\mathbf{1},\mathbf{2})$ and we will consider the brane contents and the monodromies on the $t$ and $z$ slices respectively and show that the results are consistent and match the anomaly cancellation condition in Section \ref{sec:IVs*III}.}
\label{tab:summary}
\end{table}

\subsection{Type $I_n$, $n \geq 2$}\label{sec:In}

For type $I_n$ fiber, using a Tate model we have:
\begin{align}
	\Delta = z^n(a_{10}^4P+\cO(z))
\end{align}
where $a_{10}\in\cO(-K_B)$, which is already of normal crossing type. From the multiplicity of vanishing of $(f,g,\Delta)$ at $z = P = 0$ there is an $A_{n-1}\rightarrow A_n$ enhancement,  and we expect a single hypermultiplet in $\mathbf{n}$ of $SU(n)$. At $z = a_1 = 0$ there is an $A_{n-1}\rightarrow D_n$ enhancement, and we expect a single hypermultiplet in $\mathbf{\Lambda}^2$ of $SU(n)$. These are well-known facts from perturbative string theory and can be deduced by either smoothing the singular geometry via small resolutions or deforming the 7-brane configuration \cite{Katz:1996xe, Grassi:2013kha}. We will show how to obtain the correct spectrum without resolution or deformation. We note that our method for the computation of the matter works whether there exists a smooth or terminal Calabi-Yau minimal resolution, that is, our method is insensitive to the presence of terminal and not smooth singularities, as we can see from comparing with \cite{GrassiWeigand2}. Therefore we will exclude the case $I_1$ as the singularity at the matter point is terminal and generically admits no crepant resolution though in this case one can obtain the correct (uncharged) localized matter spectrum at the matter point as expected from the perturbative limit. We will elaborate this point further in Section \ref{sec:neutralhypers}. By constructing a type $I_{ns}$ fiber using the Tate model, and then putting it into Weierstrass form, one can read off the roots of Eq.~\ref{eqn:rhs} near $z = 0$, which take the form
\begin{align}
	x_1 = \frac{1}{12}a_{10}^2,\ x_2 = -\frac{1}{6}a_{10}^2,\ x_3 = \frac{1}{12}a_{10}^2.
\end{align}
We see that $x_1$ and $x_3$ coincide therefore $x_1$-$x_3$ can be fixed as the vanishing $\pi_1$ cycle along $z = 0$. Along $L_G$ we have $z = \epsilon e^{i\theta}$ and it is easy to show that on $L_G$ we have $x_1-x_3\sim z^{\frac{n}{2}}$. This means that upon traversing $z = 0$ once, the roots $x_1$ and $x_3$ have swapped $n$ times and that corresponds exactly to the monodromy $M_{L_G} = \begin{pmatrix}
1 & n \\
0 & 1
\end{pmatrix}$. We therefore have $S_G = \{\pi_1,\dots,\pi_1\}$ where there are $n$ $\pi_1$'s. It now remains to determine $S_R$ and the monodromy of $L_R$ for the two types of enhancement.

\subsubsection{Matter in $\mathbf{n}$}\label{sec:In_fund_matter}

In this case $L_R$ is centered at $P = 0$ and is parameterized by $\delta e^{i\phi}$. It is easy to show that along $L_R$, $x_1-x_3\sim \delta^{\frac{1}{2}}e^{\frac{i\phi}{2}}$. Therefore upon traversing $L_R$, $x_1$ and $x_3$ swapped once hence the monodromy is $M_{L_R} = \begin{pmatrix}
1 & 1 \\
0 & 1
\end{pmatrix}$. Therefore in this case $S_R = \{\pi_1\}$. We have $S_G + S_R = \{\pi_1,\dots,\pi_1,\pi_1\}$ where there are $n+1$ $\pi_1$'s at $z = P = 0$ hence there is a branching rule at the point of enhancement:
\begin{align}
	SU(n+1)&\rightarrow SU(n):\nonumber\\
	\text{adj}_{SU(n+1)}&\rightarrow\text{adj}_{SU(n)} + \mathbf{n} + \overline{\mathbf{n}} + \mathbf{1}
\end{align}
There are $n$ junctions with $a(J) = (1,0)$ that give rise to $\mathbf{n}$ and $n$ junctions with $a(J) = (-1,0)$ that give rise to $\overline{\mathbf{n}}$.  To illustrate our method we compute the junctions explicitly using an $SU(5)$ model and the result is listed in Table \ref{tab:junctions_fund}.
\begin{table}[h]
\centering
	\begin{tabular}{c|c|c}
		Junction & $SU(n)$ charge & $a(J)$ \\
		\hline
        $(1, 0, 0, 0, 0, -1)$ & $(0,0,0,1)$ & $(1,0)$ \\
        $(0, 1, 0, 0, 0, -1)$ & $(0,0,1,-1)$ & $(1,0)$ \\
        $(0, 0, 1, 0, 0, -1)$ & $(0,1,-1,0)$ & $(1,0)$ \\
        $(0, 0, 0, 1, 0, -1)$ & $(1,-1,0,0)$ & $(1,0))$ \\
        $(0, 0, 0, 0, 1, -1)$ & $(-1,0,0,0)$ & $(1,0)$ \\
 		\end{tabular}
\caption{The 5 junctions corresponding to $\mathbf{5}$ of $SU(5)$ near the point of enhancement $z = P = 0$. The middle column is the charge of the state under the Cartan $U(1)$'s of $SU(5)$. $a(J)$ is computed with respect to $S_G$. The 5 junctions corresponding to $\overline{\mathbf{5}}$ are the orientation reversed junctions of the ones listed here.}
\label{tab:junctions_fund}
\end{table}

The relevant matter-point monodromy matrices are:
\begin{align}
	M_{1} = \begin{pmatrix}
1 & n+1 \\
0 & 1
\end{pmatrix}, \qquad M_{2} = \begin{pmatrix}
1 & n-1 \\
0 & 1
\end{pmatrix}.
\end{align}
We see that the $a(J) = (1,0)$ junctions and the $a(J) = (-1,0)$ junctions are preserved by $M_1$ and $M_2$ therefore no reduction is induced by the monodromy. Hence we obtain a hypermultiplet in $\mathbf{n}$ of $SU(n)$ at $z = P = 0$ as required by 6D anomaly cancellation.

\subsubsection{Matter in $\mathbf{\Lambda}^2$}\label{sec:ast_matter}

This case can be worked out easily using perturbative string techniques but is subtle using our method. But we will see that the string junction description nicely reproduces the spectrum required by 6D anomaly cancellation.

In this case $L_R$ is centered at $a_{10} = 0$. Note that there are six solutions to $z = \tilde\Delta = 0$ with respect to $a_{10}$ but only four of them vanish when $z = 0$ hence these are the four points inside $L_R$. We assume that $L_R = L_{R_4}\circ L_{R_3}\circ L_{R_2}\circ L_{R_1}$ where $L_{R_i}$ is the loop enclosing $a_{10,i}$ inside $L_R$.  It is easy to see that the geometric monodromy around $L_R$ is an overall $4\pi$ rotation of $x_1,x_2,x_3$ with $x_1$-$x_3$ rotated by $-4\pi$, i.e., swapped four times in an opposite direction with respect to the overall $4\pi$ rotation of the three roots. At first sight it might seem hard to derive what monodromy matrix it corresponds to but we can either geometrically show what the vanishing cycles are of the four relevant $a_{10}$'s hence derive the monodromy, or algebraically compare it with the well known result from perturbative string theory. We will see in this example that these two approaches nicely match each other.

We can choose an arbitrary type $I_{ns}$ Tate model, for example, $I_{5s}$ with $SU(5)$ gauge algebra. The discriminant locus is of the form:
\begin{align}
	\Delta = z^5\tilde\Delta = z^5(a_{10}^4P+a_{10}^2K_1z+K_2z^2+\cO(z^3))
\end{align}
and $f$ and $g$ are of the form:
\begin{align}
	f &= a_{10}Q_f(z) + \cO(z^2),\nonumber \\
	g &= a_{10}Q_g(z) + \cO(z^3)\nonumber
\end{align}
where $Q_f$ is linear in $z$ and $Q_g$ is quadratic in $z$.

Solving $\Delta = 0$ with respect to $z$ there are $5+k$ solutions where $k$ is the order of $\tilde\Delta$ in $z$. Besides the five roots at $z = 0$ that correspond to the five gauge branes, two of the remaining roots $z = z_A$ and $z = z_B$ approach $z = 0$ in the limit $a_{10} = 0$. It is easy to see from the forms of $f$, $g$ and $\Delta$ that the $a_{10} = 0$ limit is the limit where the fiber becomes type $I_{1s}^*$ and the gauge algebra becomes $SO(10)$. In this sense the $SU(5)$ gauge theory along $z = 0$ is a Higgsing of the $SO(10)$ gauge theory by a deformation parameterized by $a_{10}$. To match the result in perturbative string theory we must have that the branes at $z = 0$ have vanishing cycles $(1,0)$ and the two branes along $z = z_A$ and $z = z_B$ together form an $O7$ plane in the $a_{10} = 0$ limit. In the spirit of Section \ref{sec:reptheory} the $SL(2,\mathbb{Z})$ monodromy associated with the $O7$ plane stays the same under deformation. When $a_{10}$ is small, the $SO(10)$ theory is Higgsed and the branes $z_A$ and $z_B$ are separated from $z = 0$ therefore we are able to choose a loop $L_{A+B}$ around both $z = z_A$ and $z = z_B$ and the monodromy corresponding to this loop is the monodromy that corresponds to that of an $O7$ plane. It is now easy to see that upon going along $L_{A+B}$, the geometric motion of $x_1,x_2,x_3$ is an overall $2\pi$ rotation together with a $-2\pi$ rotation of $x_1$-$x_3$. We see that the geometric monodromy corresponding to $L_R$ is twice the geometric monodromy corresponding to an $O7$ plane. We know that
\begin{align}
	M_{O7} = \begin{pmatrix}
		-1 & 4 \\
		0 & -1
	\end{pmatrix}.
\end{align}
Therefore, according to Eq. \ref{eq:geomrep} we have:
\begin{align}
	M_{L_R} = \begin{pmatrix}
		1 & -8 \\
		0 & 1
	\end{pmatrix}.
\end{align}

We also need to deduce the vanishing cycles corresponding to each of the four $a_{10}$'s inside $L_R$ to read off the brane content on the disk $D$. This can be determined easily by observing that the geometric motion of $x_1,x_2,x_3$ around the loops $L_{R_2}\circ L_{R_1}$ and $L_{R_4}\circ L_{R_3}$ are both that of the loop around an $O7$ plane. Therefore we have
\begin{align}
	M_{L_{R_2}\circ L_{R_1}} = M_{L_{R_4}\circ L_{R_3}} = \begin{pmatrix}
		-1 & 4 \\
		0 & -1
	\end{pmatrix}.
\end{align}
Geometrically it is a bit tricky to read off the vanishing cycles corresponding to $a_{10,2}$ and $a_{10,3}$ but it is pretty clear that the vanishing cycles associated with $a_{10,1}$ and $a_{10,4}$ are $\pi_2$ and $\pi_3$ by approaching these two points respectively. Therefore, according to Eq. \ref{eq:geomrep} we must have
\begin{align}
	M_{L_{R_2}}M_{\pi_2} = M_{\pi_3}M_{L_{R_3}} = \begin{pmatrix}
		-1 & 4 \\
		0 & -1
	\end{pmatrix}.
\end{align}
Therefore it can be determined that
\begin{align}
	M_{L_{R_2}} = \begin{pmatrix}
		2 & 1\\
		-1 & 0
	\end{pmatrix},\ M_{L_{R_3}} = \begin{pmatrix}
		-1 & 4\\
		-1 & 3
	\end{pmatrix}.
\end{align}
Hence we can read off the vanishing cycles, $v_{a_{10,2}} = (1,-1)$, $v_{a_{10,3}} = (2,1)$, and we have
\begin{align}
	S_G + S_R = \{\pi_1,\pi_1,\dots,\pi_1,\pi_3,\pi_\beta,\pi_\alpha,\pi_2\}
\end{align}

We can also read off the vanishing cycles directly from the geometry instead of using Eq. \ref{eq:geomrep}. This requires looking at the elliptic fiber $E$ which is a double cover of the $x$-plane branched at $x_1,x_2,x_3$ and the point at infinity instead of only looking at the $x$-plane.
\begin{figure}[h]
	\centering
	\begin{tikzpicture}
		\filldraw[black] (-1,0) circle (2pt) node[anchor=east] {$x_2$};
		\filldraw[black] (1,1) circle (2pt) node[anchor=west] {$x_1$};
		\filldraw[black] (1,-1) circle (2pt) node[anchor=west] {$x_3$};
		\filldraw[black] (0,-.5)  node[below] {$v_1$};
		\filldraw[black] (1.8,2.5)  node[below] {$v_2$};
		\filldraw[black] (0.2,0.85) node[] {$\pi_{1}$};
		\draw[red, thick] (1,-1) -- (-1,0);
		\draw[black, thick] (1,1) -- (-1,0);
		\useasboundingbox (-1,0) rectangle (2.5,2.5);
		\draw[red, thick] (-1,0) .. controls (1,4) and (3.75,1.5) .. (1,-1);
	\end{tikzpicture}
	\caption{Ramification points away from infinity of the generic fiber. The black line denotes the choice of a branch cut, $v_{1}$ and $v_{2}$ correspond to two vanishing cycles with endpoints $x_{1}$ and $x_{2}$.}
	\label{fig:vcycle}
\end{figure}
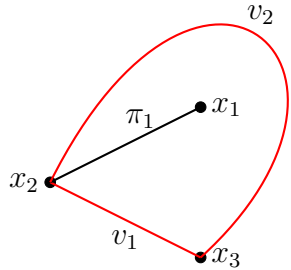

In order to read off the vanishing cycles corresponding to $M_{L_{R_{2}}}$ and $M_{L_{R_{3}}}$, we note that these correspond to the collapse of $v_{2}$ as pictured in figure~\ref{fig:vcycle} instead of $v_{1}$. Letting $\pi_{1}$ denote the vanishing cycle encircling the branch cut, we observe that the cycle $v_{2}$ simply corresponds to the image of a picard lefschetz monodromy acting on $v_{1}$ under a clockwise rotation of $\pi_{1}$ by $2\pi$. Thus, letting $T_{\pi_{1}}$ denote the matrix corresponding to the picard lefschetz monodromy, we conclude that $v_{2} = T_{\pi_{1}}^{-2}v_{1}$. Performing the calculation in the canonical basis, we have $T_{\pi_{1}}^{-2} = \begin{pmatrix}
	1 & -2 \\
	0 & 1
\end{pmatrix}$. Acting on the cycle $v_{1} = (-1,-1)$, we find that this maps to $v_{2} = (1,-1)$. A completely analogous calculation maps the cycle $(0,1)$ to $(2,1)$ via a counter-clockwise rotation, and we conclude.

Using the branes $S_G+S_R$ we can search for junctions that give rise to matters in this system. We will use $SU(5)$ as a concrete example and we list the junctions corresponding to the highest weight state of $\Lambda^2$ in the following table:
\begin{table}[h]
\centering
	\begin{tabular}{c|c|c}
		Junction & $SU(n)$ charge & $a(J)$ \\
		\hline
        $(1, 1, 0, 0, 0, -1, 1, -2, 2)$ & $(0,0,1,0)$ & $(2,0)$ \\
        $(1, 1, 0, 0, 0, 0, 0, -1, 1)$ & $(0,0,1,0)$ & $(2,0)$ \\
        $(1, 1, 0, 0, 0, 1, -1, 0, 0)$ & $(0,0,1,0)$ & $(2,0)$ \\
        $(1, 1, 0, 0, 0, 2, -2, 1, -1)$ & $(0,0,1,0)$ & $(2,0)$
 		\end{tabular}
\caption{The 4 junctions corresponding to the highest weight state of $\mathbf{\Lambda}^2$ of $SU(5)$ near the point of enhancement $z = a_{10} = 0$. The middle column is the charge of the state under the Cartan $U(1)$'s of $SU(5)$. $a(J)$ is computed with respect to $S_G$.}
\label{tab:junctions_ast}
\end{table}

The monodromy matrices $M_L$ and $M_{\bar{L}}$ are both of the form $\begin{pmatrix}
	1 & k \\
	0 & 1
\end{pmatrix}$ therefore the junctions with $a(J) = (\pm 2,0)$ are invariant under their action. Hence it seems that there are 4 $\mathbf{\Lambda^2}$'s and 4 $\mathbf{\overline{\Lambda^2}}$'s to which the junctions correspond are the orientation reversed junctions of the ones that give rise to $\mathbf{\Lambda^2}$.

\begin{figure}[h]
	\centering
	\begin{tikzpicture}
		\filldraw[red] (-2,1) circle (2pt) node[anchor=east] {$\pi_\alpha$};
		\filldraw[red] (-2,-1) circle (2pt) node[anchor=east] {$\pi_\beta$};
		\filldraw[red] (-1,1) circle (2pt) node[anchor=west] {$\pi_2$};
		\filldraw[red] (-1,-1) circle (2pt) node[anchor=west] {$\pi_3$};
		\filldraw[blue] (1,0) circle (2pt) node[anchor=west] {$SU(5)$};
		\draw[dashed] (-0.5,0) ellipse (3cm and 2cm);
	\end{tikzpicture}\caption{On disk $D$ there are four $I_1$'s and the $SU(5)$ gauge branes. The dashed curve denotes a closed $(n,0)$ string that encloses the system $S_G+S_R$.}
	\label{fig:astmatter}
\end{figure}
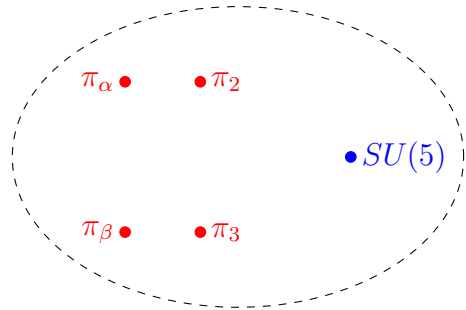

But in this case there is a subtlety that will also appear later when we discuss the type $IV^*$ and type $III^*$ models. We see that due to the form of the monodromy, there are closed eigen-strings that encloses the brane system $S_G+S_R$. Closed strings that carry charge $(n,0)$ can enclose $S_G+S_R$ since $(n,0)$ is invariant under the total monodromy $M_L$. This configuration is sketched in Figure \ref{fig:astmatter}.

Here we see that the codimension 2 monodromy is:
\begin{align}
	M = \begin{pmatrix}
		1 & k \\
		0 & 1
	\end{pmatrix}.
\end{align}
The only charge vector $a(J)$ with eigenvalue 1 under $M$ is $a(J) = (1,0)$ and indeed this is the only eigenvector of $M$. Since the eigenvalue is 1, a string emitting charge $a(J) = (1,0)$ can close back onto itself therefore becoming a closed string. We may call this an ``closed eigen-string".

In particular we can choose a closed eigen-string that carries charge $a(J) = (1,0)$. Via Hanany-Witten moves we can show that this closed eigen-string is equivalent to the junction
\begin{align}
	Q_C = \pm(0,0,0,0,0,1,-1,1,-1)
\end{align}
where the sign is determined by the direction of the charge running in the closed string. We choose $J = (1,1,0,0,0,1,-1,0,0)$ then we can see that all the other three junctions in Table \ref{tab:junctions_ast} are of the form $J + nQ_C$ where $n\in\mathbb{Z}$. Moreover, the intersection matrix corresponding to $S_G+S_R$ is:
\begin{align}
	I = \begin{pmatrix}
 -1 & 0 & 0 & 0 & 0 & \frac{1}{2} & \frac{1}{2} & -\frac{1}{2} & -\frac{1}{2} \\
 0 & -1 & 0 & 0 & 0 & \frac{1}{2} & \frac{1}{2} & -\frac{1}{2} & -\frac{1}{2} \\
 0 & 0 & -1 & 0 & 0 & \frac{1}{2} & \frac{1}{2} & -\frac{1}{2} & -\frac{1}{2} \\
 0 & 0 & 0 & -1 & 0 & \frac{1}{2} & \frac{1}{2} & -\frac{1}{2} & -\frac{1}{2} \\
 0 & 0 & 0 & 0 & -1 & \frac{1}{2} & \frac{1}{2} & -\frac{1}{2} & -\frac{1}{2} \\
 \frac{1}{2} & \frac{1}{2} & \frac{1}{2} & \frac{1}{2} & \frac{1}{2} & -1 & -1 &
   -\frac{1}{2} & \frac{1}{2} \\
 \frac{1}{2} & \frac{1}{2} & \frac{1}{2} & \frac{1}{2} & \frac{1}{2} & -1 & -1 &
   -\frac{3}{2} & -\frac{1}{2} \\
 -\frac{1}{2} & -\frac{1}{2} & -\frac{1}{2} & -\frac{1}{2} & -\frac{1}{2} & -\frac{1}{2}
   & -\frac{3}{2} & -1 & -1 \\
 -\frac{1}{2} & -\frac{1}{2} & -\frac{1}{2} & -\frac{1}{2} & -\frac{1}{2} & \frac{1}{2}
   & -\frac{1}{2} & -1 & -1 \\
\end{pmatrix}.
\end{align}
To obtain $a(J) = (2,0)$ junctions with respect to $S_G$, the following condition has to be satisfied:
\begin{align}
	n_1(0,1)+n_2(2,1)+n_3(1,-1)+n_4(-1,-1) = (-2,0)
\end{align}
where $n_i\in\mathbb{N}$. We then have:
\begin{align}
	n_3 = \frac{1}{2}(n_1-n_2) - 1,\ n_4 = \frac{1}{2}(n_1+3n_2)+1.
\end{align}
We then need to find $J = (J_G, n_1, n_2, n_3, n_4)$ such that $(J,J) = -2$ where $J_G$ is a junction with $a(J_G) = (2,0)$ and $(J_G, J_G) = -2$ with respect to $S_G$. Solving $(J,J) = -2$ we have $n_2 = -n_1$ or $n_2 = 1-n_1$. We see immediately that when $n_2 = 1-n_1$, $n_3$ and $n_4$ are not integers therefore we must have $n_2 = -n_1$. Hence the junctions that give rise to $\mathbf{\Lambda^2}$ are all of the form:
\begin{align}
	J = (J_G,n_1,-n_1,n_1-1,1-n_1) = (J_G,1,-1,0,0)+(n_1-1)Q_C.
\end{align}

We have shown that $Q_C$ is indeed a closed string so that all the junctions except those of the form $J = (J_G,1,-1,0,0)$ are superpositions of a junction that stretches between $S_G$ and $S_R$ and a junction that is in fact a closed string. Therefore such states are not the matters that are localized at the point of enhancement and we are led to the conclusion that only one hypermultiplet in $\mathbf{\Lambda^2}$ is the true localized matter at the point of enhancement. This matches the result that is required by 6D anomaly cancellation.

\subsection{Type $I_n^*$}

In this case $S_G = \{\pi_1,\pi_3,\pi_1,\pi_3,\pi_1,\pi_3,\pi_1,\dots,\pi_1\}$ where there are in total $n+6$ branes. When $n$ is even we have:
\begin{equation}
	\Delta = z^{n+6}(a_{21}^2(a_{4,2+\frac{n}{2}}^2 + P) + O(z)),
\end{equation}
while when $n$ is odd we have:
\begin{equation}
	\Delta = z^{n+6}(a_{21}^3a_{3,\frac{n+3}{2}}^2 + O(z)).
\end{equation}

We consider first the locus where there is matter in $\textbf{vect}$ as required by anomaly cancellation where $a_{4,2+\frac{n}{2}} = z = 0$ (or $a_{3,\frac{n+3}{2}} = z = 0$). The situation is exactly the same for the cases that $n = \text{even}$ or the cases that $n = \text{odd}$ therefore for simplicity we consider the case $n = 0$. Here we require the gauge group to be $SO(8)$ where we have:
\begin{equation}
	\Delta = z^6(a_{21}^2(a_{21}^2 + P_1)(a_{21}^2 + P_2) + O(z)).
\end{equation}
The matters are localized at $z = a_{21}^2 = 0$, $z = a_{21}^2 + P_1 = 0$ and $z = a_{21}^2 + P_2 = 0$.

At $z = a_{21}^2 = 0$ the brane content is
\begin{equation}
	S_G + S_R = \{\pi_1,\pi_3,\pi_1,\pi_3,\pi_1,\pi_3,\pi_1,\pi_1\}
\end{equation}
and we have:
\begin{equation}
	M_L = \begin{pmatrix}
		-1 & -2 \\
		0 & -1
	\end{pmatrix}, \qquad
	M_{\bar{L}} = \begin{pmatrix}
		-1 & 2 \\
		0 & -1
	\end{pmatrix}. \label{eq:I0*_monodromy}
\end{equation}
There are four sets of junctions in the representation $\mathbf{8_{\text{v}}}$ out of which the junctions with highest weights are:
\begin{table}[h]
\centering
	\begin{tabular}{c|c|c}
		Junction & $SO(8)$ charge & $a(J)$ \\
		\hline
        $( 1, 0, 0, 0, 0, 0, -1, 0 )$ & $(0,0,0,1)$ & $( 1 , 0 )$ \\
        $( 1, 0, 0, 0, 0, 0, 0, -1 )$ & $(0,0,0,1)$ & $( 1 , 0 )$ \\
        $( 1, 1, -1, 0, -1, -1, 0, 1 )$ & $(0,0,0,1)$ & $( -1 , 0 )$ \\
        $( 1, 1, -1, 0, -1, -1, 1, 0 )$ & $(0,0,0,1)$ & $( -1 , 0 )$
 		\end{tabular}
\caption{The four junctions with highest weight of $\mathbf{8_{\text{v}}}$ under $SO(8)$.}
\label{tab:junctions_I0*}
\end{table}

\smallskip

Upon the reductions given by Eq. \ref{eq:I0*_monodromy} we see that the matter content is $\mathbf{8_{\text{v}}} + \mathbf{8_{\text{v}}}$ which is the hypermultiplet in $\mathbf{8_{\text{v}}}$ of $SO(8)$ as required by 6D anomaly cancellation.

The above lines of computations also apply to the other cases in the $I_n^*$ series. For an $I_n^*$ fiber at the locus $z = a_{4,2+\frac{n}{2}} = 0$ the brane content is
\begin{equation}
	S_G + S_R = \{\pi_1,\pi_3,\pi_1,\pi_3,\pi_1,\pi_3,\pi_1, \dots, \pi_1, \pi_1, \pi_1\}
\end{equation}
where in $S_G + S_R$ there are in total $n + 5$ 7-branes with $\pi_1$ vanishing cycle and the monodromies are:
\begin{equation}
	M_L = \begin{pmatrix}
		-1 & n-2 \\
		0 & -1
	\end{pmatrix}, \qquad
	M_{\bar{L}} = \begin{pmatrix}
		-1 & n+2 \\
		0 & -1
	\end{pmatrix}. \label{eq:In*_monodromy}
\end{equation}

There are four sets of junctions in $\textbf{vect}$ of $SO(2n + 8)$ out of which there are two with $a(J) = (1,0)$ and the other two with $a(J) = (-1,0)$. Upon the reductions given by Eq. \ref{eq:In*_monodromy} the matter content is a hypermultiplet in $\textbf{vect}$ of $SO(2n+8)$ that meets the requirement of 6D anomaly cancellation.

The other locus $z = a_{21} = 0$ does not always exist for the type of elliptic fibration  with $I_n^*$  fibers considered here. Recall in fact  that for $I_n^*$ the Weierstrass model is:
\begin{align*}
	f &= -3a_{21}^2z^2 + Fa_{21}z^3 + O(z^4), \\
	g &= 2a_{21}^3z^3 + Ga_{21}^2z^4+ O(z^5), \\
	\Delta &= z^{6+n}(a_{21}^2Q + O(z)).
\end{align*}
When $n\geq 4$ the order of vanishing of $(f,g,\Delta)$ at $z = a_{21} = 0$ exceeds $(4,6,12)$ and this situation is beyond the scope of this paper.

When the locus $z = a_{21} = 0$ exists in $I_n^*$ fibration, there will be matters in $\textbf{spin}$ of $SO(2n + 8)$. We again use $SO(8)$ as an example. In this example the matters are localized at $z = a_{21}^2 + P_1 = 0$ and $z = a_{21}^2 + P_2 = 0$. At $z = a_{21}^2 + P_1 = 0$ the brane content is:
\begin{equation}
	S_G + S_R = \{\pi_1,\pi_3,\pi_1,\pi_3,\pi_1,\pi_3,\pi_2,\pi_2\}
\end{equation}
and we have:
\begin{equation}
	M_L = \begin{pmatrix}
		1 & -2 \\
		2 & -3
	\end{pmatrix}, \qquad
	M_{\bar{L}} = \begin{pmatrix}
		-3 & 2 \\
		-2 & 1
	\end{pmatrix}. \label{eq:I0*s_monodromy}
\end{equation}
There are four sets of junctions in the representation $\mathbf{8_{\text{s}}}$ out of which the junctions with highest weights are:
\begin{table}[h]
\centering
	\begin{tabular}{c|c|c}
		Junction & $SO(8)$ charge & $a(J)$ \\
		\hline
        $( 0, 1, -1, -1, 0, -1, -1, 0 )$ & $(0,0,1,0)$ & $( -1 , -1 )$ \\
        $( 0, 1, -1, -1, 0, -1, 0, -1 )$ & $(0,0,1,0)$ & $( -1 , -1 )$ \\
        $( 1, 1, 0, 0, 0, 0, 0, 1 )$ & $(0,0,1,0)$ & $( 1 , 1 )$ \\
        $( 1, 1, 0, 0, 0, 0, 1, 0 )$ & $(0,0,1,0)$ & $( 1 , 1 )$
 		\end{tabular}
\caption{The four junctions with highest weight of $\mathbf{8_{\text{s}}}$ under $SO(8)$.}
\label{tab:junctions_I0*s}
\end{table}

At $z = a_{21}^2 + P_3 = 0$ the brane content is:
\begin{equation}
	S_G + S_R = \{\pi_1,\pi_3,\pi_1,\pi_3,\pi_1,\pi_3,\pi_3,\pi_3\}
\end{equation}
and we have:
\begin{equation}
	M_L = \begin{pmatrix}
		-1 & 0 \\
		2 & -1
	\end{pmatrix}, \qquad
	M_{\bar{L}} = \begin{pmatrix}
		-1 & 0 \\
		-2 & -1
	\end{pmatrix}. \label{eq:I0*c_monodromy}
\end{equation}
There are four sets of junctions in the representation $\mathbf{8_{\text{c}}}$ out of which the junctions with highest weights are:
\begin{table}[h]
\centering
	\begin{tabular}{c|c|c}
		Junction & $SO(8)$ charge & $a(J)$ \\
		\hline
        $( 0, 0, 0, 0, 0, -1, 0, 1 )$ & $(1,0,0,0)$ & $( 0 , -1 )$ \\
        $( 0, 0, 0, 0, 0, -1, 1, 0 )$ & $(1,0,0,0)$ & $( 0 , -1 )$ \\
        $( 1, 1, 0, 1, -1, -1, -1, 0 )$ & $(1,0,0,0)$ & $( 0 , 1 )$ \\
        $( 11, 1, 0, 1, -1, -1, 0, -1 )$ & $(1,0,0,0)$ & $( 0 , 1 )$
 		\end{tabular}
\caption{The four junctions with highest weight of $\mathbf{8_{\text{c}}}$ under $SO(8)$.}
\label{tab:junctions_I0*c}
\end{table}

In both of the above two cases, it is easy to see that upon the reduction of the monodromies given by Eq. \ref{eq:I0*s_monodromy} or Eq. \ref{eq:I0*c_monodromy}, the matter content is a hypermultiplet in either a hypermultiplet in $\mathbf{8_{\text{s}}}$ or a hypermultiplet in $\mathbf{8_{\text{c}}}$ which satisfies the requirement of 6D anomaly cancellation.

\subsection{Type $III$} \label{sec:III}

The Weierstrass model we are using for type $III$ is:
\begin{align}
	\begin{split}
		f &= tz, \\
		g &= z^2
	\end{split}
\end{align}
of which the discriminant locus is:
\begin{equation}
	\Delta = z^3(27z+4t^3).
\end{equation}
The 7-branes near the point of enhancement $z = t = 0$ are $S_G = \{\pi_1,\pi_3,\pi_2\}$ and $S_R = \{\pi_1,\pi_3,\pi_2\}$. Therefore there is a branching rule at the point of enhancement:
\begin{align}
	SO(8)&\rightarrow SU(2):\nonumber\\
	\mathbf{28}&\rightarrow \mathbf{3}+4\times(\mathbf{2}+\mathbf{2})+9\times\mathbf{1}
\end{align}
The junctions at the point of enhancement are listed in Table. \ref{tab:junctions_III}.
\begin{table}[h]
\centering
	\begin{tabular}{c|c|c}
		Junction & $SU(2)$ charge & $a(J)$ \\
		\hline
        $( -1 , -1 , 0 , 0 , 0 , -1 )$ & -1 & $( -1 , -1 )$ \\
        $( -1 , -1 , 0 , 1 , 1 , 0 )$ & -1 & $( -1 , -1 )$ \\
        $( -1 , 0 , 0 , 0 , -1 , -1 )$ & -1 & $( -1 , 0 )$ \\
        $( -1 , 0 , 0 , 1 , 0 , 0 )$ & -1 & $( -1 , 0 )$ \\
        $( 0 , -1 , -1 , -1 , 0 , 0 )$ & -1 & $( 1 , 0 )$ \\
        $( 0 , -1 , -1 , 0 , 1 , 1 )$ & -1 & $( 1 , 0 )$ \\
        $( 0 , 0 , -1 , -1 , -1 , 0 )$ & -1 & $( 1 , 1 )$ \\
        $( 0 , 0 , -1 , 0 , 0 , 1 )$ & -1 & $( 1 , 1 )$ \\
        $( 0 , 0 , 1 , 0 , 0 , -1 )$ & 1 & $( -1 , -1 )$ \\
        $( 0 , 0 , 1 , 1 , 1 , 0 )$ & 1 & $( -1 , -1 )$ \\
        $( 0 , 1 , 1 , 0 , -1 , -1 )$ & 1 & $( -1 , 0 )$ \\
        $( 0 , 1 , 1 , 1 , 0 , 0 )$ & 1 & $( -1 , 0 )$ \\
        $( 1 , 0 , 0 , -1 , 0 , 0 )$ & 1 & $( 1 , 0 )$ \\
        $( 1 , 0 , 0 , 0 , 1 , 1 )$ & 1 & $( 1 , 0 )$ \\
        $( 1 , 1 , 0 , -1 , -1 , 0 )$ & 1 & $( 1 , 1 )$ \\
        $( 1 , 1 , 0 , 0 , 0 , 1 )$ & 1 & $( 1 , 1 )$ \\
 		\end{tabular}
\caption{The 16 junctions near the point of enhancement $III\rightarrow I_0^*$. The middle column is the charge of the state under the Cartan $U(1)$ of $SU(2)$. $a(J)$ is computed with respect to the first three 7-branes which are the gauge 7-branes.}
\label{tab:junctions_III}
\end{table}

The monodromy matrix $M_L$ is that of type $I_0^*$, $M_L = \begin{pmatrix}
-1 & 0 \\
0 & -1
\end{pmatrix}$ while it is clear that $M_{\bar{L}}$ is trivial. The orbits of the asymptotic charges under $M_L$ are:
\begin{align}
	& \text{Orbit 1}:\ (1,0)\rightarrow(-1,0) \\
	& \text{Orbit 2}:\ (1,1)\rightarrow(-1,-1)
\end{align}
%
%

Both Orbit 1 and Orbit 2 give $\mathbf{2}$. Out of the 8 $\mathbf{2}$'s obtained from the branching rule, there are two of them with $a(J)=(1,0)$, two of them with $a(J)=(-1,0)$, two of them with $a(J)=(1,1)$ and two of them with $a(J)=(-1,-1)$. The first two sets of junction are both on Orbit 1, and the latter two sets of junctions are both on Orbit 2. So after identifying the states via monodromy, there are 4 $\mathbf{2}$'s left which become two $\mathbf{2}$ full hypers as required by 6D anomaly cancellation.

\subsection{Type $IV_s$} \label{sec:IVs}

The Weierstrass model we are using for type $IV_s$  is:
\begin{align}
	\begin{split}
		f &= z^2, \\
		g &= t^2z^2
	\end{split}
\end{align}
of which the discriminant locus is:
\begin{equation}
	\Delta = z^4(4z^2+27t^4).
\end{equation}
The 7-branes near the point of enhancement $z = t = 0$ are $S_G = \{\pi_1,\pi_3,\pi_1,\pi_3\}$ and $S_R = \{\pi_1,\pi_3,\pi_1,\pi_3\}$. Therefore there is a branching rule at the point of enhancement:
\begin{align}
	E_6&\rightarrow SU(3):\\
	\mathbf{78}&\rightarrow \mathbf{8}+9\times(\mathbf{3}+\mathbf{\overline{3}})+16\times\mathbf{1}
\end{align}

The monodromy matrix $M_L$ is that of type $IV^*$, $M_L = \begin{pmatrix}
0 & -1 \\
1 & -1
\end{pmatrix}$ while it is clear that $M_{\bar{L}}$ is trivial. The orbits of the asymptotic charges under $M_L$ are:
\begin{align}
	& \text{Orbit 1}:\ (1,0)\rightarrow(0,1)\rightarrow(-1,-1)\rightarrow(1,0) \\
	& \text{Orbit 2}:\ (-1,0)\rightarrow(0,-1)\rightarrow(1,1)\rightarrow(-1,0)
\end{align}

The junctions at the point of enhancement are listed in Table. \ref{tab:junctions_IVs}.
\begin{center}
	\begin{longtable}[h]{c|c|c}
		Junction & $SU(3)$ charge & $a(J)$ \\
		\hline
        $( -1 , -1 , 0 , 0 , 0 , 0 , 1 , 1 )$ & $( -1 , 0 )$ & $( -1 , -1 )$ \\
        $( -1 , -1 , 0 , 0 , 1 , 0 , 0 , 1 )$ & $( -1 , 0 )$ & $( -1 , -1 )$ \\
        $( -1 , -1 , 0 , 0 , 1 , 1 , 0 , 0 )$ & $( -1 , 0 )$ & $( -1 , -1 )$ \\
        $( -1 , -1 , 1 , 0 , 0 , 0 , 0 , 1 )$ & $( 0 , -1 )$ & $( 0 , -1 )$ \\
        $( -1 , -1 , 1 , 0 , 0 , 1 , 0 , 0 )$ & $( 0 , -1 )$ & $( 0 , -1 )$ \\
        $( -1 , -1 , 1 , 0 , 1 , 1 , -1 , 0 )$ & $( 0 , -1 )$ & $( 0 , -1 )$ \\
        $( -1 , 0 , 0 , -1 , 0 , 0 , 1 , 1 )$ & $( 1 , -1 )$ & $( -1 , -1 )$ \\
        $( -1 , 0 , 0 , -1 , 1 , 0 , 0 , 1 )$ & $( 1 , -1 )$ & $( -1 , -1 )$ \\
        $( -1 , 0 , 0 , -1 , 1 , 1 , 0 , 0 )$ & $( 1 , -1 )$ & $( -1 , -1 )$ \\
        $( -1 , 0 , 0 , 0 , 0 , -1 , 1 , 1 )$ & $( 0 , -1 )$ & $( -1 , 0 )$ \\
        $( -1 , 0 , 0 , 0 , 0 , 0 , 1 , 0 )$ & $( 0 , -1 )$ & $( -1 , 0 )$ \\
        $( -1 , 0 , 0 , 0 , 1 , 0 , 0 , 0 )$ & $( 0 , -1 )$ & $( -1 , 0 )$ \\
        $( 0 , -1 , 0 , 0 , 0 , 0 , 0 , 1 )$ & $( -1 , 1 )$ & $( 0 , -1 )$ \\
        $( 0 , -1 , 0 , 0 , 0 , 1 , 0 , 0 )$ & $( -1 , 1 )$ & $( 0 , -1 )$ \\
        $( 0 , -1 , 0 , 0 , 1 , 1 , -1 , 0 )$ & $( -1 , 1 )$ & $( 0 , -1 )$ \\
        $( 0 , -1 , 1 , 1 , -1 , 0 , 0 , 0 )$ & $( -1 , 0 )$ & $( 1 , 0 )$ \\
        $( 0 , -1 , 1 , 1 , 0 , 0 , -1 , 0 )$ & $( -1 , 0 )$ & $( 1 , 0 )$ \\
        $( 0 , -1 , 1 , 1 , 0 , 1 , -1 , -1 )$ & $( -1 , 0 )$ & $( 1 , 0 )$ \\
        $( 0 , 0 , -1 , -1 , 0 , 0 , 1 , 1 )$ & $( 0 , 1 )$ & $( -1 , -1 )$ \\
        $( 0 , 0 , -1 , -1 , 1 , 0 , 0 , 1 )$ & $( 0 , 1 )$ & $( -1 , -1 )$ \\
        $( 0 , 0 , -1 , -1 , 1 , 1 , 0 , 0 )$ & $( 0 , 1 )$ & $( -1 , -1 )$ \\
        $( 0 , 0 , -1 , 0 , 0 , -1 , 1 , 1 )$ & $( -1 , 1 )$ & $( -1 , 0 )$ \\
        $( 0 , 0 , -1 , 0 , 0 , 0 , 1 , 0 )$ & $( -1 , 1 )$ & $( -1 , 0 )$ \\
        $( 0 , 0 , -1 , 0 , 1 , 0 , 0 , 0 )$ & $( -1 , 1 )$ & $( -1 , 0 )$ \\
        $( 0 , 0 , 0 , -1 , 0 , 0 , 0 , 1 )$ & $( 1 , 0 )$ & $( 0 , -1 )$ \\
        $( 0 , 0 , 0 , -1 , 0 , 1 , 0 , 0 )$ & $( 1 , 0 )$ & $( 0 , -1 )$ \\
        $( 0 , 0 , 0 , -1 , 1 , 1 , -1 , 0 )$ & $( 1 , 0 )$ & $( 0 , -1 )$ \\
        $( 0 , 0 , 0 , 1 , -1 , -1 , 1 , 0 )$ & $( -1 , 0 )$ & $( 0 , 1 )$ \\
        $( 0 , 0 , 0 , 1 , 0 , -1 , 0 , 0 )$ & $( -1 , 0 )$ & $( 0 , 1 )$ \\
        $( 0 , 0 , 0 , 1 , 0 , 0 , 0 , -1 )$ & $( -1 , 0 )$ & $( 0 , 1 )$ \\
        $( 0 , 0 , 1 , 0 , -1 , 0 , 0 , 0 )$ & $( 1 , -1 )$ & $( 1 , 0 )$ \\
        $( 0 , 0 , 1 , 0 , 0 , 0 , -1 , 0 )$ & $( 1 , -1 )$ & $( 1 , 0 )$ \\
        $( 0 , 0 , 1 , 0 , 0 , 1 , -1 , -1 )$ & $( 1 , -1 )$ & $( 1 , 0 )$ \\
        $( 0 , 0 , 1 , 1 , -1 , -1 , 0 , 0 )$ & $( 0 , -1 )$ & $( 1 , 1 )$ \\
        $( 0 , 0 , 1 , 1 , -1 , 0 , 0 , -1 )$ & $( 0 , -1 )$ & $( 1 , 1 )$ \\
        $( 0 , 0 , 1 , 1 , 0 , 0 , -1 , -1 )$ & $( 0 , -1 )$ & $( 1 , 1 )$ \\
        $( 0 , 1 , -1 , -1 , 0 , -1 , 1 , 1 )$ & $( 1 , 0 )$ & $( -1 , 0 )$ \\
        $( 0 , 1 , -1 , -1 , 0 , 0 , 1 , 0 )$ & $( 1 , 0 )$ & $( -1 , 0 )$ \\
        $( 0 , 1 , -1 , -1 , 1 , 0 , 0 , 0 )$ & $( 1 , 0 )$ & $( -1 , 0 )$ \\
        $( 0 , 1 , 0 , 0 , -1 , -1 , 1 , 0 )$ & $( 1 , -1 )$ & $( 0 , 1 )$ \\
        $( 0 , 1 , 0 , 0 , 0 , -1 , 0 , 0 )$ & $( 1 , -1 )$ & $( 0 , 1 )$ \\
        $( 0 , 1 , 0 , 0 , 0 , 0 , 0 , -1 )$ & $( 1 , -1 )$ & $( 0 , 1 )$ \\
        $( 1 , 0 , 0 , 0 , -1 , 0 , 0 , 0 )$ & $( 0 , 1 )$ & $( 1 , 0 )$ \\
        $( 1 , 0 , 0 , 0 , 0 , 0 , -1 , 0 )$ & $( 0 , 1 )$ & $( 1 , 0 )$ \\
        $( 1 , 0 , 0 , 0 , 0 , 1 , -1 , -1 )$ & $( 0 , 1 )$ & $( 1 , 0 )$ \\
        $( 1 , 0 , 0 , 1 , -1 , -1 , 0 , 0 )$ & $( -1 , 1 )$ & $( 1 , 1 )$ \\
        $( 1 , 0 , 0 , 1 , -1 , 0 , 0 , -1 )$ & $( -1 , 1 )$ & $( 1 , 1 )$ \\
        $( 1 , 0 , 0 , 1 , 0 , 0 , -1 , -1 )$ & $( -1 , 1 )$ & $( 1 , 1 )$ \\
        $( 1 , 1 , -1 , 0 , -1 , -1 , 1 , 0 )$ & $( 0 , 1 )$ & $( 0 , 1 )$ \\
        $( 1 , 1 , -1 , 0 , 0 , -1 , 0 , 0 )$ & $( 0 , 1 )$ & $( 0 , 1 )$ \\
        $( 1 , 1 , -1 , 0 , 0 , 0 , 0 , -1 )$ & $( 0 , 1 )$ & $( 0 , 1 )$ \\
        $( 1 , 1 , 0 , 0 , -1 , -1 , 0 , 0 )$ & $( 1 , 0 )$ & $( 1 , 1 )$ \\
        $( 1 , 1 , 0 , 0 , -1 , 0 , 0 , -1 )$ & $( 1 , 0 )$ & $( 1 , 1 )$ \\
        $( 1 , 1 , 0 , 0 , 0 , 0 , -1 , -1 )$ & $( 1 , 0 )$ & $( 1 , 1 )$ \\
		\caption{The 54 junctions near the point of enhancement $IV_s\rightarrow IV^*$. The middle column is the charge of the state under the Cartan $U(1)^2$ of $SU(3)$. $a(J)$ is computed with respect to the first four 7-branes which are the gauge 7-branes.}\label{tab:junctions_IVs}\\
	\end{longtable}
\end{center}

Orbit 1 gives $\mathbf{3}$ and Orbit 2 gives $\mathbf{\overline{3}}$. We first consider the 9 $\mathbf{3}$'s. There are three of them with $a(J)=(1,0)$, three of them with $a(J)=(0,1)$ and three of them with $a(J)=(-1,-1)$. We see that these three sets are all on Orbit 1 so after identifying states via monodromy, there are 3 $\mathbf{3}$'s left. Next we consider the 9 $\mathbf{\overline{3}}$'s. There are three of them with $a(J)=(1,1)$, three of them with $a(J)=(0,-1)$ and three of them with $a(J)=(-1,0)$. We see that these three sets are all on Orbit 2 so after identifying states via monodromy, there are 3 $\mathbf{\overline{3}}$'s left. In total there are 3 $(\mathbf{3}+\mathbf{\overline{3}})$'s left which are the three full hypers required by 6D anomaly cancellation.

\subsection{Type $IV^*_s$}\label{sec:IV*}

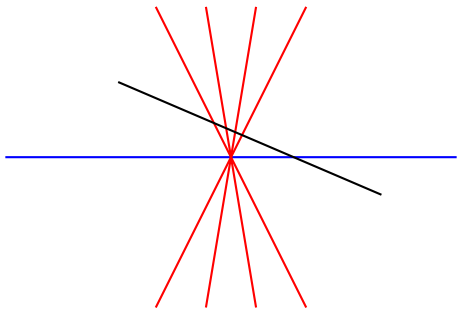
\begin{figure}[h]
	\centering
	\begin{tikzpicture}
		\draw[blue, thick] (-3,0) -- (3,0);
		\draw[red, thick] (-1,2) -- (1,-2);
		\draw[red, thick] (-1/3,2) -- (1/3,-2);
		\draw[red, thick] (1/3,2) -- (-1/3,-2);
		\draw[red, thick] (1,2) -- (-1,-2);
		\draw[black, thick] (-1.5,1) -- (2,-0.5);
	\end{tikzpicture}\caption{The gauge 7-branes of $E_6$ algebra are on top of each other and denoted by the red line. The four red lines denote the four $I_1$'s which are obtained from solving the equation $\tilde{\Delta} = 0$ with respect to $g_4$. The black line segment denotes a disk $D$ that intersects the 12 7-branes.}
	\label{fig:IVstar_codim2}
\end{figure}

The Weierstrass model we are using for type $IV^*_s$ is:
\begin{align}
	\begin{split}
		f &= z^3, \\
		g &= t^2 z^4
	\end{split}
\end{align}
of which the discriminant locus is:
\begin{equation}
	\Delta = z^8(4z+27t^4).
\end{equation}
The 7-branes near the point of enhancement $z = t = 0$ are
\begin{equation*}
	S_G = \{\pi_1,\pi_3,\pi_1,\pi_3,\pi_1,\pi_3,\pi_1,\pi_3\},\ S_R = \{\pi_1,\pi_3,\pi_1,\pi_3\}.
\end{equation*}
We see that there are 12 7-branes at codimension 2. The first eight 7-branes are the $E_6$ gauge branes and the last four 7-branes are the $I_1$'s associated with the solutions of $\tilde{\Delta} = 0$ with respect to $g_4$. Near the enhancement point $z = g_4 = 0$ the geometry is sketched in Figure \ref{fig:IVstar_codim2}.

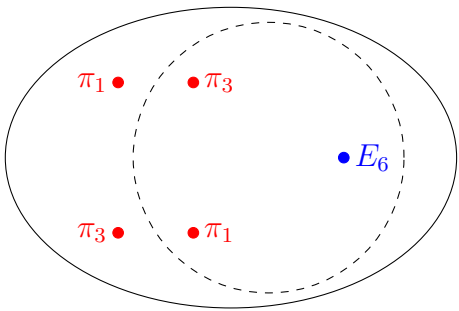
\begin{figure}[h]
	\centering
	\begin{tikzpicture}
		\filldraw[red] (-2,1) circle (2pt) node[anchor=east] {$\pi_1$};
		\filldraw[red] (-2,-1) circle (2pt) node[anchor=east] {$\pi_3$};
		\filldraw[red] (-1,1) circle (2pt) node[anchor=west] {$\pi_3$};
		\filldraw[red] (-1,-1) circle (2pt) node[anchor=west] {$\pi_1$};
		\filldraw[blue] (1,0) circle (2pt) node[anchor=west] {$E_6$};
		\draw (-0.5,0) ellipse (3cm and 2cm);
		\draw[dashed] (0,0) circle (1.8cm);
	\end{tikzpicture}\caption{On disk $D$ there are four $I_1$'s and the $E_6$ gauge branes. The 10 branes enclosed by the dashed circle are $\{\pi_1,\pi_3,\pi_1,\pi_3,\pi_1,\pi_3,\pi_1,\pi_3,\pi_1,\pi_3\}$ which are the brane content of an $E_8$ algebra}.
	\label{fig:IVstar_disk}
\end{figure}

We choose a disk $D$ that intersects the 12 7-branes near the enhancement point. On $D$ the branes can be organized into a set of 10 7-branes that realizes an $E_8$ algebra and two extra branes $\pi_1$ and $\pi_3$ as shown in Figure \ref{fig:IVstar_disk}.

\begin{figure}[h]
	\centering
	\begin{tikzpicture}
		\filldraw[red] (-2,0) circle (2pt) node[anchor=east] {$\pi_1$};
		\filldraw[red] (2,0) circle (2pt) node[anchor=east] {$\pi_3$};
		\filldraw[black] (0,0) circle (4pt) node[anchor=west] {$E_8$};
		\draw[dashed] (0,0) ellipse (4cm and 2.5cm);
		\draw[dashed] (1,0) circle (1.8cm);
		\draw[dashed] (-1,0) circle (1.8cm);
		\node at (-1.5,1.3) {$l_3$};
		\node at (1.5,-1.3) {$l_1$};
		\node at (0,2.2) {$l_2$};
	\end{tikzpicture}\caption{The $E_8$ branes obtained from the grouping together the $E_6$ gauge branes and the extra $\pi_1$ and $\pi_3$ branes are denoted by the big black point. $l_1$ is a junction that loops around the $E_8$ branes and a $\pi_3$ brane with charge $a(l_1) = (1,0)$. $l_3$ is a junction that loops around the $E_8$ branes and a $\pi_1$ brane with charge $a(l_3) = (0,1)$. $l_2$ is a junction that loops around all the 12 7-branes with charge $a(l_2) = (1,1)$.}
	\label{fig:dP9}
\end{figure}
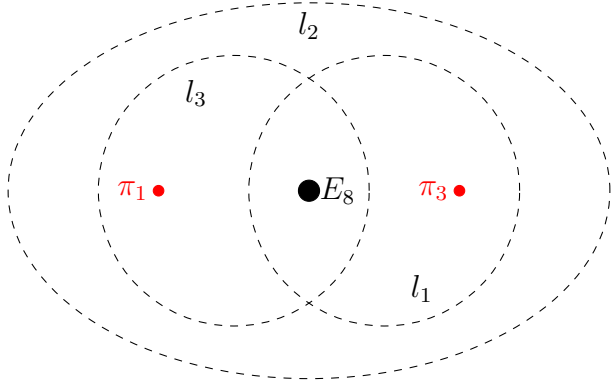

The branes in Figure \ref{fig:IVstar_codim2} can then be grouped to form the configuration shown in Figure \ref{fig:dP9}.
We can see that because of their asymptotic charges, $l_1$ and $l_3$ can actually be deformed to junctions that loop around all the 12 7-branes via trivial Hanany-Witten moves since $l_1$ can be pulled across the branch cut of $\pi_1$ without creating a new prong and $l_3$ can be pulled across the branch cut of $\pi_3$ without creating a new prong. Via Hanany-Witten move we can show that $l_1$ is equivalent to the junction $Q_1 = (1, 1, 0, 1, -1, 0, -1, -1, 0, -1, 1, 0)$, $l_3$ is equivalent to the junction $Q_3 = (1, 0, 1, 1, 0, 1, -1, 0, -1, -1, 0, -1)$ and $l_2$ is equivalent to the junction $Q_2 = (2, 1, 1, 2, -1, 1, -2, -1, -1, -2, 1, -1)$. We can see that $Q_2 = Q_1 + Q_3$. Actually, all the junctions $l_L$ that loop around all the 12 7-branes of this system is a superposition of $l_1$ and $l_3$ and its charge $Q_L$ is a linear combination of $Q_1$ and $Q_3$.

The intersection matrix of this system with 12 7-branes is:
\begin{align}
	I = \begin{psmallmatrix}
-1 & \frac{1}{2} & 0 & \frac{1}{2} & 0 & \frac{1}{2} & 0 & \frac{1}{2} & 0 & \frac{1}{2} & 0 & \frac{1}{2}\\
\frac{1}{2} & -1 & -\frac{1}{2} & 0 & -\frac{1}{2} & 0 & -\frac{1}{2} & 0 & -\frac{1}{2} & 0 & -\frac{1}{2} & 0\\
0 & -\frac{1}{2} & -1 & \frac{1}{2} & 0 & \frac{1}{2} & 0 & \frac{1}{2} & 0 & \frac{1}{2} & 0 & \frac{1}{2}\\
\frac{1}{2} & 0 & \frac{1}{2} & -1 & -\frac{1}{2} & 0 & -\frac{1}{2} & 0 & -\frac{1}{2} & 0 & -\frac{1}{2} & 0\\
0 & -\frac{1}{2} & 0 & -\frac{1}{2} & -1 & \frac{1}{2} & 0 & \frac{1}{2} & 0 & \frac{1}{2} & 0 & \frac{1}{2}\\
\frac{1}{2} & 0 & \frac{1}{2} & 0 & \frac{1}{2} & -1 & -\frac{1}{2} & 0 & -\frac{1}{2} & 0 & -\frac{1}{2} & 0\\
0 & -\frac{1}{2} & 0 & -\frac{1}{2} & 0 & -\frac{1}{2} & -1 & \frac{1}{2} & 0 & \frac{1}{2} & 0 & \frac{1}{2}\\
\frac{1}{2} & 0 & \frac{1}{2} & 0 & \frac{1}{2} & 0 & \frac{1}{2} & -1 & -\frac{1}{2} & 0 & -\frac{1}{2} & 0\\
0 & -\frac{1}{2} & 0 & -\frac{1}{2} & 0 & -\frac{1}{2} & 0 & -\frac{1}{2} & -1 & \frac{1}{2} & 0 & \frac{1}{2}\\
\frac{1}{2} & 0 & \frac{1}{2} & 0 & \frac{1}{2} & 0 & \frac{1}{2} & 0 & \frac{1}{2} & -1 & -\frac{1}{2} & 0\\
0 & -\frac{1}{2} & 0 & -\frac{1}{2} & 0 & -\frac{1}{2} & 0 & -\frac{1}{2} & 0 & -\frac{1}{2} & -1 & \frac{1}{2}\\\frac{1}{2} & 0 & \frac{1}{2} & 0 & \frac{1}{2} & 0 & \frac{1}{2} & 0 & \frac{1}{2} & 0 & \frac{1}{2} & -1
\end{psmallmatrix}.
\end{align}
It is easy to show that with respect to $I$ the self-intersection numbers $(l_1,l_1) = (l_2,l_2) = (l_3,l_3) = 0$. The junctions correspond to the simple roots of $E_8$ are:
\begin{align*}
	\alpha_1 &= (0, 0, 0, 1, -1, -1, 0, -1, 1, 1, 0, 0), \\
	\alpha_2 &= (0, 0, 0, 0, 0, 0, 0, 1, 0, -1, 0, 0), \\
  	\alpha_3 &= (0, 0, 0, 0, 0, 0, 1, 0, -1, 0, 0, 0), \\
  	\alpha_4 &= (0, 0, 0, 0, 1, 0, -1, 0, 0, 0, 0, 0), \\
   	\alpha_5 &= (0, 0, 1, 0, -1, 0, 0, 0, 0, 0, 0, 0), \\
   	\alpha_6 &= (0, 1, -2, -1, 0, -1, 1, 0, 1, 1, 0, 0), \\
   	\alpha_7 &= (1, -1, 1, 1, 0, 1, -1, 0, -1, -1, 0, 0), \\
  	\alpha_8 &= (0, 0, 0, 0, 0, 1, -1, -1, 1, 0, 0, 0).
\end{align*}
It is easy to show that $(a_i,l_1) = (a_i,l_2) = (a_i,l_3) = 0$. Since any root $\gamma$ of $E_8$ can be written as $\gamma = \sum_i a_i\alpha_1$ with $(\gamma,\gamma) = -2$ and any junction $l_L$ that loop around all the 12 7-branes can be written as $l_L = Al_1 + Bl_3$, we see that $(l_L,l_L) = 0$, $(\gamma,l_L) = 0$ therefore $(\gamma+nl_L,\gamma+nl_L) = -2$, where $a_i$, $A$, $B$ and $n$ are all integers. We denote by $J_n$ the states $\gamma + nl_L$. It is obvious that $a(J_n) = (0,0)$.

Moreover, $S_g=\{\alpha_1,\alpha_2,\alpha_3,\alpha_4,\alpha_5,\alpha_6,\alpha_7,\alpha_8,l_1,l_3\}$ is a set that generates all the junctions with $a(J) = (0,0)$. So that all the junctions with $a(J) = (0,0)$ and $(J,J) = -2$ are of the form $\gamma + nl_L$ since $a(l_L) = 0$ and $(l_L,\psi) = 0$, $\forall\psi\in S_g$ and the only linear combinations of $\alpha_i$'s such that $(\sum_ic_i\alpha_i,\sum_ic_i\alpha_i) = -2$  and $a(\sum_ia_i\alpha_i) = (0,0)$ are the roots of $E_8$, $\gamma$.

Now we see that there is an infinite number of states $J_n = \gamma + nl_L$ such that $a(J_n) = (0,0)$ and $(J_n,J_n) = -2$ that are graded by $n$. For $n\neq 0$, $J_n$ is a junction that is a superposition of a junction $\gamma$ that stretches between $E_6$ gauge branes and the $I_1$ locus and a junction $l_L$ that is a closed string around the 12 7-branes of the system. These states are not the matters that are localized at the enhancement point at codimension 2 on the base. To extract only the matter content at codimension 2, we need to focus on the junctions $J_0$, i.e., $\gamma$'s that stretch between the $E_6$ gauge branes and the $I_1$ locus without being superposed with a closed string $l_L$.

Therefore in this system what we actually have is an enhancement from $E_6$ to $E_8$, the string junctions can be derived via the branching rule:
\begin{align}
	E_8&\rightarrow E_6:\\
	\mathbf{248}&\rightarrow \mathbf{78}+3\times(\mathbf{27}+\mathbf{\overline{27}})+8\times\mathbf{1}
\end{align}

The relevant monodromies are:
\begin{align}
	M_{L_R} = \begin{pmatrix}
-1 & 1 \\
-1 & 0
\end{pmatrix},\ M_{L_G} = \begin{pmatrix}
0 & -1 \\
1 & -1
\end{pmatrix}.
\end{align}
Therefore we have:
\begin{align}
	M_{L} = \begin{pmatrix}
		1 & 0 \\
		0 & 1
	\end{pmatrix},\ M_{\bar{L}} = \begin{pmatrix}
		-1 & 1 \\
		-1 & 0
	\end{pmatrix}
\end{align}
We see that $M_{L}$ is trivial and the orbit of $M_{\bar{L}}$ is:
\begin{align}
	\text{Orbit}:\ \begin{matrix}
		(1,0)\\
		(-1,0)
	\end{matrix}\rightarrow\begin{matrix}
		(-1,-1)\\
		(1,1)
	\end{matrix}\rightarrow\begin{matrix}
		(0,1)\\
		(0,-1)
	\end{matrix}
\end{align}

Out of the three full hypers in $\mathbf{27}$, one of them is with $a(J) = (\pm1,0)$, one the them with $a(J) = \pm (1,1)$ and one of them with $a(J) = (0,\pm1)$. We see that all of them are on the orbit of either $M_{\bar{L}}$ so they will be identified via the monodromy and there is only one full hyper in $\mathbf{27}$ left in the spectrum which is the matter content required by 6D anomaly cancellation.

\subsection{Type $III^*$}

The Weierstrass model we are using for type $III^*$ is:
\begin{align}
	\begin{split}
		f &= tz^3, \\
		g &= z^5
	\end{split}
\end{align}
of which the discriminant locus is:
\begin{equation}
	\Delta = z^9(27z+4t^3).
\end{equation}
The 7-branes near the point of enhancement $z = t = 0$ are:
\begin{equation*}
	S_G = \{\pi_1,\pi_3,\pi_1,\pi_3,\pi_1,\pi_3,\pi_1,\pi_3,\pi_1\},\ S_R = \{\pi_3,\pi_1,\pi_3\}.
\end{equation*}
By comparing with the result in Section \ref{sec:IV*} we see that near the $z = t = 0$ the brane content $S_G+S_R$ is exactly the same as that of type $IV^*$. It is clear that the argument in Section \ref{sec:IV*} also holds in this case since it only uses the data of the brane content on the disk $D$ but not the details of the Weierstrass model. Therefore in this case the gauge algebra is also effectively enhanced to $E_8$. Hence we have the branching rule:
\begin{align}
	E_8&\rightarrow E_7\nonumber \\
	\mathbf{248}&\rightarrow\mathbf{133} + 2\times\mathbf{56} + 3.
\end{align}

The relevant monodromies are:
\begin{align}
	M_{L_R} = \begin{pmatrix}
0 & 1 \\
-1 & 0
\end{pmatrix},\ M_{L_G} = \begin{pmatrix}
0 & 1 \\
-1 & 0
\end{pmatrix}.
\end{align}
Therefore we have:
\begin{align}
	M_{L} = \begin{pmatrix}
		1 & 0 \\
		0 & 1
	\end{pmatrix},\ M_{\bar{L}} = \begin{pmatrix}
		-1 & 0 \\
		0 & -1
	\end{pmatrix}
\end{align}
We see that $M_{L}$ is trivial and the orbit of $M_{\bar{L}}$ is:
\begin{align}
	\text{Orbit}:(0,1)\rightarrow(0,-1)\rightarrow(0,1).
\end{align}

The two $\mathbf{56}$'s obtained from string junction computation are one with $a(J) = (0,1)$ and one with $a(J) = (-1,0)$. We see that they are on the orbit of $M_{\bar{L}}$ so they are identified. We are left with one $\mathbf{56}$ that forms half hypermultiplet in $\mathbf{56}$ of $E_7$. This is the matter content that is required by 6D anomaly cancellation.

\subsection{$III\times III$}

The Weierstrass model we are using for type $III\times III$ is:
\begin{align}
	\begin{split}
		f &= zt, \\
		g &= z^2t^2
	\end{split}
\end{align}
of which the discriminant locus is:
\begin{equation}
	\Delta = z^3t^3(4 + 27zt).
\end{equation}
The 7-branes near the intersection point $z = t = 0$ are:
\begin{equation*}
	S_G = \{\pi_1,\pi_3,\pi_2,\pi_1,\pi_3,\pi_2\}.
\end{equation*}
It is easy to see the branching rule is:
\begin{align*}
	SO(8)&\rightarrow SU(2)\times SU(2): \\
	\mathbf{28}&\rightarrow(\mathbf{3}, \mathbf{1}) + (\mathbf{1}, \mathbf{3}) + 4\times(\mathbf{2}, \mathbf{2}) + 6\times(\mathbf{1}, \mathbf{1}).
\end{align*}
Note that there are two sets of gauge 7-branes $S_A$ and $S_B$, therefore we label the asymptotic charge of the junctions by $a(J) = (Q_A, Q_B)$ where $Q_A$ is the asymptotic charge associated with $S_A$ and $Q_B$ the asymptotic charge associated with $S_B$.

The 4 $(\mathbf{2}, \mathbf{2})$ are respectively with asymptotic charge:
\begin{align*}
	a(J_1) &= ((1,0),(-1,0)), \\
	a(J_2) &= ((-1,0),(1,0)), \\
	a(J_3) &= ((1,1),(-1,-1)), \\
	a(J_4) &= ((-1,-1),(1,1)).
\end{align*}

We have:
\begin{equation}
	M_L = \begin{pmatrix}
		-1 & 0 \\
		0 & -1
	\end{pmatrix},\quad M_{\bar{L}} = \begin{pmatrix}
		1 & 0 \\
		0 & 1
	\end{pmatrix}.
\end{equation}
Therefore we have $J_1\rightarrow J_2$ and $J_3\rightarrow J_4$ under the monodromies $M_L$ and $M_{\bar{L}}$. This gives rise to one hypermultiplet in the bifundamental representation of $SU(2)\times SU(2)$ which matches the result of 6D anomaly cancellation.

\subsection{$IV_s\times IV_s$}

The Weierstrass model we are using for type $IV_s\times IV_s$ is:
\begin{align}
	\begin{split}
		f &= z^2t^2, \\
		g &= z^2t^2
	\end{split}
\end{align}
of which the discriminant locus is:
\begin{equation}
	\Delta = z^4t^4(27 + 4z^2t^2).
\end{equation}
The 7-branes near the intersection point $z = t = 0$ are:
\begin{equation*}
	S_G = \{\pi_1,\pi_3,\pi_1,\pi_3,\pi_1,\pi_3,\pi_1,\pi_3\}.
\end{equation*}
It is easy to see the branching rule is:
\begin{align*}
	E_6&\rightarrow SU(3)\times SU(3): \\
	\mathbf{78}&\rightarrow(\mathbf{8}, \mathbf{1}) + (\mathbf{1}, \mathbf{8}) + 3\times((\mathbf{3}, \mathbf{\overline{3}}) + (\mathbf{\overline{3}}, \mathbf{3})) + 8\times(\mathbf{1}, \mathbf{1}).
\end{align*}
Again there are two sets of gauge 7-branes $S_A$ and $S_B$, therefore we label the asymptotic charge of the junctions by $a(J) = (Q_A, Q_B)$ where $Q_A$ is the asymptotic charge associated with $S_A$ and $Q_B$ the asymptotic charge associated with $S_B$.

The 3 $(\mathbf{3}, \mathbf{\overline{3}})$ are respectively with asymptotic charge:
\begin{align*}
	a(J_1) &= ((1,0),(-1,0)), \\
	a(J_2) &= ((0,1),(0,-1)), \\
	a(J_3) &= ((-1,-1),(1,1)).
\end{align*}
And the 3 $(\mathbf{\overline{3}}, \mathbf{3})$ are respectively with asymptotic charge:
\begin{align*}
	a(J_4) &= ((-1,0),(1,0)), \\
	a(J_5) &= ((0,-1),(0,1)), \\
	a(J_6) &= ((1,1),(-1,-1)).
\end{align*}

We have:
\begin{equation}
	M_L = \begin{pmatrix}
		0 & -1 \\
		1 & -1
	\end{pmatrix},\quad M_{\bar{L}} = \begin{pmatrix}
		1 & 0 \\
		0 & 1
	\end{pmatrix}.
\end{equation}
Therefore we have $J_1\rightarrow J_2\rightarrow J_3$ and $J_4\rightarrow J_5\rightarrow J_6$ under the monodromies $M_L$ and $M_{\bar{L}}$. This gives rise to one hypermultiplet in the bifundamental representation of $SU(3)\times SU(3)$ which matches the result of 6D anomaly cancellation.

\subsection{$IV_s\times III$}\label{sec:IVs*III}

The Weierstrass model we are using for type $IV_s \times III$ is:

\begin{align}
	\begin{split}
		f &= zt^{2}, \\
		g &= z^2 t^2
	\end{split}
\end{align}
of which the discriminant locus is:
\begin{equation}
	\Delta = z^3 t^4(4t^2 + 27z)
\end{equation}

\noindent
We will consider two natural 7-brane systems corresponding to the triple intersection dictated by the discriminant locus. Consider the brane system corresponding to the t-slice:

\begin{equation*}
	S_{G_t} = \{\pi_1,\pi_3,\pi_1,\pi_3,\pi_1,\pi_3,\pi_1,\pi_3\}.
\end{equation*}

\noindent
The corresponding branching rule at the point of enhancement is:
\begin{align*}
	E_6&\rightarrow SU(3)\times SU(2): \\
	\mathbf{78}&\rightarrow(\mathbf{8}, \mathbf{1}) + (\mathbf{1}, \mathbf{3}) + 3\times((\mathbf{3}, \mathbf{2}) + (\mathbf{\overline{3}}, \mathbf{2})) + 3\times((\mathbf{3}, \mathbf{1}) + (\mathbf{\overline{3}}, \mathbf{1})) + ((\mathbf{1},\mathbf{2}) + (\mathbf{1},\mathbf{2})) + 9\times(\mathbf{1}, \mathbf{1})
\end{align*}

\noindent
The monodromy matrix of this system is identical to that of type $IV_s \times IV_s$, \begin{equation}
	M_L = \begin{pmatrix}
		0 & -1 \\
		1 & -1
	\end{pmatrix},\quad M_{\bar{L}} = \begin{pmatrix}
		1 & 0 \\
		0 & 1
	\end{pmatrix}.
\end{equation}

As in the above, there are two sets of gauge 7-branes $S_A$ and $S_B$, but a novelty of the $IV_s\times III$ model is that there is an extra $I_1$ brane (or two extra $I_1$ branes as we will see momentarily) therefore we will label the corresponding asymptotic charges of the junctions by $a(J) = (Q_A,Q_B,Q_{I_1})$ with $Q_A$ the asymptotic charge with respect to $S_A$, $Q_B$ the asymptotic charge with respect to $S_B$ and $Q_{I_1}$ with respect to the extra $I_1$.

This induces a partition of the set $S_{G_t} = \{ S_A, S_B, I_1\}$, where $S_A$ is the first 4 branes in $S_{G_t}$ corresponding to the locus $t^4 = 0$ and the set $\{S_B, I_1\}$ denote the intersection of the $SU(2)$ and $I_1$ seven-brane intersections with the $t$-slice.


We will consider only the subset of junctions charged under $S_A$ and the action of the monodromy $M$ with respect to the asymptotic charge $Q_A$. The $3$ $(\mathbf{3},\mathbf{1})$ carry the asymptotic charges:
\begin{align*}
	a(J_1) &= ((0,1)_A,(0,0)_B,(0,-1)_{I_1}), \\
	a(J_2) &= ((-1,-1)_A,(1,1)_B,(0,0)_{I_1}), \\
	a(J_3) &= ((1,0)_A,(-1,1)_B,(0,-1)_{I_1}).
\end{align*}
while the $3$ $(\mathbf{3},\mathbf{2})$ carry the asymptotic charges:
\begin{align*}
	a(J_4) &= ((0,1)_A,(0,-1)_B,(0,0)_{I_1}), \\
	a(J_5) &= ((-1,-1)_A,(1,0)_B,(0,1)_{I_1}), \\
	a(J_6) &= ((1,0)_A,(-1,0)_B,(0,0)_{I_1}).
\end{align*}

Therefore we have $J_1\rightarrow J_2\rightarrow J_3$ and $J_4\rightarrow J_5\rightarrow J_6$ under the monodromies $M_L$ and $M_{\bar{L}}$. This gives rise to one hypermultiplet in the $(\mathbf{3},\mathbf{1})$ and one hypermultiplet in the bifundamental representation of $SU(3)\times SU(2)$ which matches the result of 6D anomaly cancellation.

Instead, taking the brane system corresponding to the $z$-slice, we have the following:

\begin{equation*}
	S_{G_z} = \{\pi_1,\pi_3,\pi_1,\pi_1,\pi_3,\pi_1,\pi_3,\pi_1,\pi_3\}.
\end{equation*}
The corresponding branching rule at the point of enhancement is:
\begin{align*}
	E_7&\rightarrow SU(3)\times SU(2): \\
	\mathbf{133}&\rightarrow(\mathbf{8}, \mathbf{1}) + (\mathbf{1}, \mathbf{3}) + 4\times((\mathbf{3}, \mathbf{2}) + (\mathbf{\overline{3}}, \mathbf{2})) + 7\times((\mathbf{3}, \mathbf{1}) + (\mathbf{\overline{3}}, \mathbf{1})) + 4\times((\mathbf{1},\mathbf{2}) + (\mathbf{1},\mathbf{2})) + 16\times(\mathbf{1}, \mathbf{1})
\end{align*}
The monodromy matrix of this system is given by \begin{equation}
	M_L = \begin{pmatrix}
		0 & -1\\
		1 & 0
	\end{pmatrix},\quad M_{\bar{L}} = \begin{pmatrix}
		0 & 1 \\
		-1 & 0
	\end{pmatrix}.
\end{equation}

With notation as in the above, this set of 7-branes is naturally partitioned as $S_{G_z} = \{S_B,S_A,I_1\}$ where we now have the set $\{S_A,I_1\}$ denoting the intersection of the $SU(3)$ and $2$ $I_1$ seven-brane intersections with the $z$-slice.
The $8$ $(\mathbf{1},\mathbf{2})$ carry the asymptotic charges:
\begin{align*}
	a(J_1) &= ((0,0)_B,(0,-1)_A,(0,1)_{I_1}), \\
	a(J_2) &= ((1,2)_B,(0,-1)_A,(-1,-1)_{I_1}), \\
	a(J_3) &= ((0,0)_B,(-1,0)_A,(1,0)_{I_1}), \\
	a(J_4) &= ((1,-1)_B,(-1,0)_A,(0,1)_{I_1}), \\
	a(J_5) &= ((0,0)_B,(0,1)_A,(0,-1)_{I_1}), \\
	a(J_6) &= ((-1,-2)_B,(0,1)_A,(1,1)_{I_1}), \\
	a(J_7) &= ((0,0)_B,(1,0)_A,(-1,0)_{I_1}), \\
	a(J_8) &= ((-1,1)_B,(1,0)_A,(0,-1)_{I_1})
\end{align*}
while the $4$ $(\mathbf{3},\mathbf{2})$ carry the asymptotic charges:
\begin{align*}
	a(J_9) &= ((0,1)_B,(0,-1)_A,(0,0)_{I_1}), \\
	a(J_{10}) &= ((1,0)_B,(-1,0)_A,(0,0)_{I_1}), \\
	a(J_{11}) &= ((0,1)_B,(1,0)_A,(-1,-1)_{I_1}), \\
	a(J_{12}) &= ((-1,-1)_B,(0,1)_A,(1,0)_{I_1}), \\
\end{align*}

Therefore we have $\{J_1, J_2\} \rightarrow \{J_3, J_4\} \rightarrow \{J_5, J_6\} \rightarrow \{J_7, J_8\}$ and $J_9 \rightarrow J_{10} \rightarrow J_{11} \rightarrow J_{12}$ under the monodromies $M_L$ and $M_{\bar{L}}$. This gives rise to one hypermultiplet in the $(\mathbf{1},\mathbf{2})$ and one hypermultiplet in the bifundamental representation of $SU(3) \times SU(2)$.

Combining the results from the different slices, we find in total a charged hypermultiplet spectrum of $(\mathbf{3},\mathbf{2})+(\mathbf{3},\mathbf{1})+(\mathbf{1},\mathbf{2})$, matching the anomaly cancelling spectrum of \cite{Grassi:2014zxa}. Note that the bifundamental massless spectrum computed from the $z$-slice is the same state computed from the $t$-slice and hence is consistent with expectations.

\section{Remarks on Localized Neutral Hypermultiplets}\label{sec:neutralhypers}

In Section~\ref{sec:In} we noted that our method for the computation of the matter works whether there exists a smooth or terminal Calabi-Yau minimal resolution, that is, our method is insensitive to the presence of terminal and not smooth singularities, as we can see from comparing with~\cite{Arras:2016evy, GrassiWeigand2}. However, we remark that our method, in general, does not necessarily yield the correct uncharged matter spectrum away from the perturbative limit.

Our theory computes the localized charged matter spectrum at the intersection of seven-branes. However, localized neutral hypermultiplets may also exist \cite{Arras:2016evy}. We would like to comment on localized neutral hypermultiplets in light of our prescription. In particular, we will argue that more information must be added to our prescription to account for the appearance of localized neutral hypermultiplets.

In \cite{Arras:2016evy}, the physical significance of non-crepant resolvable singularities on an elliptically fibered Calabi-Yau threefold $X$ was investigated from the perspective of a $6d$ F-theory compactification. The central result asserts that after passing to a $\mathbb{Q}$-factorial terminal model $\hat{X} \rightarrow X$, the total number of localized, massless neutral hypermultiplets on the resulting M-theory Coulomb branch is given by the sum
\[
n^0_H = \sum_{P} m_P
\]
over singular points of $\hat{X}$ where $m_P$ is the Milnor number. Moreover, this proposal was demonstrated to be consistent with $6d$ gravitational anomaly cancellation and naturally appears as a summand of the total space of complex structure deformations of $\hat{X}$.

To demonstrate that our prescription is insensitive to the presence of localized neutral hypermultiplets, we will focus on the examples explored in \cite[Section 5.1]{Arras:2016evy}. Consider a type $III$-model with the following tunings:
\[
f = z_1 f_0, \quad g = z_1^{\mu_{g}}g_0\text{ for }\mu_g \geq 2, \quad \Delta = z_1^3(4f_0^3+ 27z_1^{2\mu_g -3}g_0^2)
\]
As demonstrated in loc cit., for $\mu_g = 2,3$, the model admits a crepant resolution, while for $\mu_g \geq 4$, the partial resolution exhibits a terminal hypersurface singularity. In particular, for $\mu_g = 4,5,7$, the isolated singularity results in the milnor numbers $\mu = 1,2,4$ respectively.

On the other hand, our prescription applied to the type $III$ model in Section~\ref{sec:III}, is insensitive to higher order terms in $z_1$ in the residual discriminant. In particular, our proposed brane content, monodromies, and calculation of junction states would yield precisely the same matter content in any of the above tuned models as in the generic case with $\mu_g = 2$. 


Therefore, we conclude that our prescription is incomplete in accounting for localized neutral matter, and we leave this avenue of investigation for future work.

\vspace{.5cm}
\noindent \textbf{Acknowledgments.}
We thank Bobby Acharya, Sakura Schafer-Nameki, Yi-Nan Wang and Timo Weigand for discussions. J.H. is supported by NSF CAREER grant PHY-1848089. The work of C.L. was partially supported by the Alfred P. Sloan Foundation Grant No. G-2019-12504 and by DOE Grant DE-SC0013607. B.S. is supported by the NSF Graduate Research Fellowship under grant DGE-1451070. J.T. is supported by a grant from the Simons Foundation (\#488569, Bobby Acharya). This material is in part based upon work supported by the NSF Grant  DMS-1440140 while A.G.  was in residence at the Mathematical Sciences Research Institute in Berkeley, California, during the Spring 2019 semester; AG gratefully acknowledges the support of a Simons Fellowship. The work of A.G. is partially supported by PRIN ``Moduli and Lie Theory''. A.G. is a member of GNSAGA of INDAM.

\newpage
\bibliographystyle{JHEP}
\bibliography{ref_tian}

\end{document}